\documentclass[manuscript]{aastex}
\usepackage{graphics}
\usepackage{graphicx}
\usepackage{sidecap}
\usepackage{url}

\slugcomment{}

\shorttitle{}
\shortauthors{}

\begin{document}



\title{Precise Stellar Radial Velocities of an M Dwarf with a Michelson Interferometer and a Medium-resolution Near-infrared Spectrograph}


\author{Philip S. Muirhead\altaffilmark{1,10,11}, 
Jerry Edelstein\altaffilmark{2}, 
David J. Erskine\altaffilmark{3}, 
Jason T. Wright\altaffilmark{4,5}, 
Matthew W. Muterspaugh\altaffilmark{6,7}, 
Kevin R. Covey\altaffilmark{1,12},  
Edward H. Wishnow\altaffilmark{2}, 
Katherine Hamren\altaffilmark{1}, 
Phillip Andelson\altaffilmark{2}, 
David Kimber\altaffilmark{2}, 
Tony Mercer\altaffilmark{8}, 
Samuel P. Halverson\altaffilmark{4}, 
Andrew Vanderburg\altaffilmark{2}, 
Daniel Mondo\altaffilmark{2}, 
Agnieszka Czeszumska\altaffilmark{9} 
and James P. Lloyd\altaffilmark{1}}


%
%


\altaffiltext{1}{Department of Astronomy, Cornell University, Ithaca, NY 14853}
\altaffiltext{2}{Space Sciences Laboratory, University of California, 7 Gauss Way, Berkeley, CA 94720-3411}
\altaffiltext{3}{Lawrence Livermore National Laboratory, Mail Stop L-487, Livermore, CA 94550}
\altaffiltext{4}{Department of Astronomy and Astrophysics, The Pennsylvania State University, University Park, PA 16801}
\altaffiltext{5}{Center for Exoplanets and Habitable Worlds, The Pennsylvania State 
University, University Park, PA 16802}

\altaffiltext{6}{Department of Mathematics and Physics, College of Arts and
Sciences, Tennessee State University, Boswell Science Hall, Nashville, TN
37209 }
\altaffiltext{7}{Tennessee State University, Center of Excellence in
Information Systems, 3500 John A. Merritt Blvd., Box No.~9501, Nashville, TN
37209-1561}
\altaffiltext{8}{Space Exploration Corporation, 1 Rocket Road, Hawthorne, CA 90250}

\altaffiltext{9}{Department of Nuclear Engineering, University of California, 4149 Etcheverry Hall, Berkeley, CA 94720-1730}
\altaffiltext{10}{NASA/NY Space Grant Graduate Fellow}
\altaffiltext{11}{Z. Carter Patten '25 Graduate Fellow}

\altaffiltext{12}{Hubble Fellow}


\begin{abstract}



Precise near-infrared radial velocimetry enables efficient detection and transit verification of low-mass extrasolar planets orbiting M dwarf hosts, which are faint for visible-wavelength radial velocity surveys.  The TripleSpec Exoplanet Discovery Instrument, or TEDI, is the combination of a variable-delay Michelson interferometer and a medium-resolution (R=2700) near-infrared spectrograph on the Palomar 200" Hale Telescope.  We used TEDI to monitor GJ 699, a nearby mid-M dwarf, over 11 nights spread across 3 months.  Analysis of 106 independent observations reveals a root-mean-square precision of less than 37 m $\rm s^{-1}$ for 5 minutes of integration time.  This performance is within a factor of 2 of our expected photon-limited precision.  We further decompose the residuals into a 33 m $\rm s^{-1}$ white noise component, and a 15 m $\rm s^{-1}$ systematic noise component, which we identify as likely due to contamination by telluric absorption lines.  With further development this technique holds promise for broad implementation on medium-resolution near-infrared spectrographs to search for low-mass exoplanets orbiting M dwarfs, and to verify low-mass transit candidates.






\end{abstract}





\section{Introduction}

M-dwarf stars are highly compelling targets for extrasolar planet surveys  \citep{Gaidos2007,Lunine2009,Charbonneau2009b}.  As host stars for transiting exoplanets, M dwarfs' low masses, radii and luminosities provide opportunities to broadly characterize low-mass exoplanets.  Consider that an exoplanet with the same mass, radius and incident stellar flux as the Earth orbiting a 0.21 $\rm M_\odot$, 0.27 $\rm R_\odot$, 3170 K, M5V star generates a 0.69 m $\rm s^{-1}$ semi-amplitude radial velocity signal on the host, has a 0.1\% transit depth, and a 0.01\% mid-infrared secondary eclipse depth.  These signals are nearly within reach of current ground and space-based astronomical instrumentation, enabling measurement of the exoplanet's orbit, mass, radius, average density and surface gravity, as well as the planet's atmospheric temperature, the role of a greenhouse effect and day/night temperature contrasts.  Compare this to an Earth-Sun analog system, which produces a 0.09 m $\rm s^{-1}$ radial velocity semi-amplitude, a 0.008\% optical transit depth, a 0.0004\% mid-IR secondary eclipse depth.  The lower signals from Earth twins orbiting Sun-like stars limit characterization to a measurement of the exoplanet radius with space-based photometry.  Current radial velocity capabilities and transit timing techniques can put limits on the masses of Earth-like exoplanets discovered to transit Sun-like stars.   However, accurately measuring their masses and surface temperatures will require broad progress in instrumentation.  Simply stated, Earth-like exoplanets orbiting M dwarfs will be much easier to characterize.



Unfortunately, the advantages M dwarf hosts provide for characterizing low-mass exoplanets come with the challenges of initial detection.  M dwarfs are upwards of 5 magnitudes fainter in $M_V$ than Sun-like stars, significantly reducing the efficiency of visible-wavelength radial velocity planet surveys.  Being intrinsically fainter, fewer bright M dwarfs occupy a single field on the sky, which reduces the efficiency of single-field transit surveys.  This is true despite the fact that M dwarfs dominate stellar populations \citep{Chabrier2003}.  Combining the $M_V$ luminosity function for single main-sequence field stars \citep{Wielen1983} with the distance modulus, one can calculate the relative number of single main-sequence stars in a magnitude-limited wide-field transit survey.  Comparing the number of single, main sequence stars with $M_V$ from 2.5 to 8.5, corresponding to F, G and K dwarfs, to the number with $M_V$ from 8.5 to 18.5, corresponding to M-type dwarfs, one finds that F, G and K dwarfs outnumber M dwarfs in magnitude-limited wide-field survey by more than 1000 to 1.  To overcome this limitation, the MEarth transit survey is individually targeting bright M dwarfs, rather than staring at a single field \citep{Nutzman2008} at the expense of continuous coverage.  




Despite the challenges, both visible-wavelength radial velocity surveys and transit surveys have had success with M dwarfs.  As of now, 27 planets have been detected around 21 M dwarfs, and two are known to transit their host stars: GJ 1214 b and GJ 436 b.  All hosts have a spectral type earlier than M5.  GJ 1214 b was initially detected by the MEarth transit survey, then confirmed by radial velocity measurements \citep{Charbonneau2009}, and GJ 436 b was initially discovered by a radial velocity survey \citep{Butler2004} with the transit detection occurring later \citep{Gillon2007}.  At this stage it is unclear whether the majority of future M dwarf transiting planets will be detected first by transit surveys then followed up with radial velocity measurements, or vice-versa, given that both individually target M dwarfs.  However, the Kepler Mission is a sensitive wide-field transit survey observing approximately 2500 M dwarfs brighter than V = 14.  If low-mass planets are common around M dwarfs, Kepler will potentially detect dozens of transiting candidates around M dwarfs which are bright enough for follow up radial velocity measurements.  Recent statistical analysis of Kepler planet candidates indicates that short period planets are, in fact, more common around low-mass dwarfs than high mass dwarfs \citep{Howard2011}, although this analysis did not include stars later than M0.  Measuring precise radial velocities of dozens of faint M dwarf hosts will prove a difficult challenge.  The effort to detect and confirm more transiting terrestrial exoplanets can be accelerated by improving radial velocity precision at near-infrared wavelengths.  M dwarf spectral energy distributions peak in the near infrared, and the many absorption lines from molecular transitions in near-infrared M dwarf spectra provide rich structure for measuring radial velocity signals \citep{Jones2009}.  Precise near-infrared radial velocities will increase the efficiency of exoplanet detection and transit verification around early M dwarfs, and will enable detection and transit verification of exoplanets orbiting M dwarfs later than M5, which have yet to be discovered.

Precise radial velocity measurements of M dwarfs must contend with activity-related radial velocity ``jitter,'' introduced by temperature inhomogeneities such as spots or flares on the rotating photosphere \citep{Saar1997}.  G- and K-type stars show a correlation between radial velocity jitter and Ca II emission, an indicator of surface activity \citep{Wright2005}.  H$\alpha$ emission is a typical indicator of surface activity on M dwarfs, and has been shown to correlate with Ca II emission \citep{Walkowicz2009}.  Using spectra of M dwarfs in the Sloan Digital Sky Survey, \citet{West2004} found the fraction of M dwarfs with H$\alpha$ emission rises steeply with later spectral type, from less than 5\% for M0 dwarfs, to over 70\% for M8 dwarfs \citep{West2004}.  This could present a challenge for precise radial velocimetry of mid-to-late M dwarfs.

However, less than 10 m $\rm s^-1$ of visible-wavelength radial velocity performance has been achieved on early M dwarfs which show H$\alpha$ emission \citep{Endl2003, Reiners2009}.  And, radial velocity jitter is expected to be significantly reduced at near-infrared wavelengths, because temperature inhomogenieties on the stellar surface have lower flux contrast \citep{Barnes2010}.  For this reason, precise near-infrared radial velocities also provide an important tool for verifying planetary candidates identified around active stars \citep{Prato2008}.  For example, visible-wavelength radial velocity measurements of the young, active star TW Hydra indicated the presence of a planet \citep{Setiawan2008}, but infrared radial velocity measurements did not \citep{Huelamo2008}, leading to the conclusion that the visible-wavelength Doppler signal was likely due to rotating spots. 


Measuring precise near-infrared radial velocities is challenging because of the lack of available high resolution spectrographs, the difficulty of calibration, and telluric interference.  Several groups have had success using the Earth's atmospheric lines to calibrate high resolution spectrographs, similar to the iodine cell technique: \citet{Blake2010} achieved 50 m $\rm s^{-1}$ of precision on an M dwarf over several years with the NIRSPEC spectrograph (R=25000) on the Keck II telescope, and \citet{Figueira2010} achieved 6 m $\rm s^{-1}$ on an Sun-like star over one week with the CRIRES spectrograph (R=100000) on the European Southern Observatory's Very Large Telescope.  So far, the best precision has been achieved by calibrating CRIRES with an ammonia gas cell, where \citet{Bean2010} achieved 5 m $\rm s^{-1}$ of long term precision on an M dwarf.

High resolution near-infrared spectroscopy is becoming an important technique for M dwarf exoplanet science, but in order to significantly impact the field it must be competitive with visible-wavelength spectroscopy on M dwarfs.  Consider GJ 1214, a V=15.1 M4.5 dwarf on which HARPS achieved 5 m $\rm s^{-1}$ of 1$\sigma$ radial velocity uncertainty with a 40 minute integration time \citep{Charbonneau2009}.  Scaling the CRIRES result to GJ 1214, the ammonia-cell technique would achieve 15 m/s of 1$\sigma$ radial velocity uncertainty with the same exposure time as HARPS.  This is because current high-resolution near-infrared spectrographs suffer from limited simultaneous bandwidth, with NIRSPEC covering 400 nm at one time and CRIRES covering 80 nm.  To overcome this, several groups have proposed or are developing high-resolution near-infrared echelle spectrographs with large simultaneous bandwidth, such as the Precision Radial Velocity Spectrograph \citep{Jones2008}, the Habitable-Zone Planet Finder \citep{Mahadevan2010, Ramsey2008}, and CARMENES \citep{Quirrenbach2010}.



High precision velocimetry and large simultaneous bandwidth can also be obtained by introducing a Michelson interferometer into the optical path between a telescope and an existing {\it medium}-resolution spectrograph, a method known as externally dispersed interferometry \citep{Erskine2003}, dispersed fixed delay interferometry \citep{Ge2006, Vaneyken2010} and also dispersed Fourier transform spectroscopy \citep{Hajian2007, Behr2009}.  The interferometer multiplies the stellar spectrum by a sinusoidal transmission comb before being dispersed by the spectrograph, which creates a moir\'e fringe pattern highly sensitive to Doppler shifts.  Medium-resolution near-infrared spectrographs are becoming widely available on 4-to-6 meter-class telescopes \citep[e.g.][]{Wilson2004,Simcoe2010} because of their smaller size and complexity compared to high-resolution spectrographs.  The availability of medium-resolution near-infrared spectrographs makes externally dispersed interferometry a promising technique for broad implementation to search for these most interesting low-mass exoplanets.

In this paper we present results from the TripleSpec Exoplanet Discovery Instrument, or TEDI, designed to measure precise near-infrared radial velocities of nearby M dwarfs.  TEDI is the combination of a Michelson interferometer and TripleSpec, a facility near-infrared spectrograph on the Palomar 200" Hale Telescope that simultaneously covers 1.00 to 2.46 $\rm \mu m$ at a resolution of 2700.

\section{Theory}\label{theory}

The theory behind externally dispersed interferometry or dispersed Fourier transform spectroscopy is described in several previous papers \citep[e.g.][]{Erskine2003, Hajian2007, Vaneyken2010}. These treatments are often specific to the instruments involved, and do not account for some of the effects that we encounter with TEDI. Here we provide a theoretical accounting of the TEDI data product and an explanation of how that product is converted into a measured change in radial velocity.

\subsection{The TEDI Data Product}\label{tedi_data_product}

 Given wavenumber $\nu = 1 / \lambda$, an intrinsic stellar spectrum $S_\nu$, spectrograph line-spread function $R_\nu$, and an interferometer optical path difference of $\tau$, an individual EDI spectrum, $I_{\nu,\tau}$, can be described as:

\begin{equation}
I_{\nu,\tau} = [S_\nu ( 1 + \cos(2\pi\tau\nu) )] * R_\nu
\end{equation}

\noindent where $*$ indicates a convolution and $( 1 + \cos(2\pi\tau\nu) )$ represents the effect of the sinusoidal transmission comb introduced by the interferometer {\it before} convolution with the spectrograph line-spread function.  The spectrograph line-spread function represents the broadening of spectral lines introduced by the spectrograph.  In order to get moir\'e fringes, the delay $\tau$ must be slightly varied by $\Delta\tau$ to modulate the resultant spectra.  This can be done by either positioning one of the interferometer mirrors such that different delays appear perpendicular to the dispersion direction of the spectrograph \citep[e.g.][]{Erskine2003, Zhao2009}, or by actively moving one of the interferometer mirrors with a piezo actuator and taking an individual spectrum at each position.  With TEDI, we use the latter approach, moving the mirror to keep the interferometer modulation within a pixel rather than across several pixels.  This dramatically reduces the effects of pixel-to-pixel calibration errors, such as effects from poor flat-fielding and/or background subtraction.


If we rewrite $\tau$ as a {\it bulk delay} plus a small {\it phase shift} ($\tau_0 + \Delta\tau$) and assume that $1/\Delta\tau$ is large compared to a resolution element of the spectrograph, then we can remove $\Delta\tau$ from the convolution integral.  In TEDI, $1/\Delta\tau$ is 3 orders of magnitude larger than a resolution element.  Applying this and rearranging Equation 1 using trigonometric identities, we can express an individual TEDI spectrum as a function of the bulk delay and phase shift:

\begin{equation}
I_{\nu,\tau_0,\Delta\tau} = A_\nu + B_\nu\cos(2\pi\Delta\tau\nu) - C_\nu\sin(2\pi\Delta\tau\nu) 
\end{equation}

\noindent where:

\begin{eqnarray}
A_\nu = S_\nu * R_\nu \\ 
B_\nu = [S_\nu \cos(2\pi\tau_0\nu)]*R_\nu \\
C_\nu = [S_\nu \sin(2\pi\tau_0\nu)]*R_\nu
\end{eqnarray}

\noindent $A_\nu$ is the spectrum of the star at the native-resolution of the spectrograph, referred to as the ``conventional'' spectrum.  $B_\nu$ and $C_\nu$ describe the moir\'e fringes that contain the signal from the high-resolution stellar features.  By slightly changing the delay of the interferometer by steps of $\Delta\tau$, and taking spectra at each step, we can fit $A_\nu$, $B_\nu$ and $C_\nu$ to the modulation at each pixel according to Equation 2.  We can then construct the {\it complex visibility}:

\begin{equation}
B_\nu - iC_\nu = [S_\nu e^{-i2\pi\tau_0\nu}] * R_\nu
\end{equation}

\noindent The complex visibility contains a real and imaginary component which can also be described by a phase, $\phi_\nu = \arctan{(B_\nu/C_\nu)}$, and a visibility, $V_\nu = \sqrt{B_\nu^2 + C_\nu^2}$. Under a small Doppler shift of the stellar spectrum $S_\nu \rightarrow S_{\nu + \Delta\nu}$, where $\Delta\nu={\Delta RV \over c}\nu$, and assuming that  $\Delta\nu$ and $1/\tau_0$ are small compared to a resolution element of the spectrograph, $\Delta\nu$ can be transferred to the exponent by treating the convolution as a Fourier transform integration and applying the Fourier shift theorem.  In this case, the Doppler-shifted {\it epoch} complex visibility, $B^1_\nu - iC^1_\nu$, is related to an unshifted {\it template} complex visibility, $B^0_\nu - iC^0_\nu$, by:


\begin{equation}
B^1_\nu - iC^1_\nu = [ B^0_\nu - iC^0_\nu ]e^{-i2\pi\tau_0\Delta\nu}
\end{equation}
 
\noindent Thus, a small Doppler shift causes a change of the phase of the complex visibility.  Since all wavelengths expect the same radial velocity shift, the change in phase versus wavelength is expected to follow a simple curve:

\begin{equation}\label{delta_phase}
\Delta\phi =  \rm 2\pi\tau_0\Delta\nu = 2\pi\tau_0(\Delta RV) \nu / c
 \end{equation}
 
\noindent where $\rm \Delta RV$ is the change in the radial velocity of the star.  However, the motion of the telescope relative to the barycenter of the Solar System introduces large Doppler shifts on the order of 10 km $\rm s^{-1}$ on most stars, which is a significant fraction of a TripleSpec pixel.  In this case, the convolution cannot be treated as Fourier transform integration and the Fourier shift theorem is not applicable.  To accurately account for a large change in the radial velocity of a star, we define a complex line-spread function as $\tilde{R}_\nu = e^{-i2\pi\tau_0(\nu)} R_{\nu}$.  We can now rewrite Equation 6 as:

\begin{equation}
B_\nu - iC_\nu = e^{i2\pi\tau_0\nu} [ S_\nu * \tilde{R}_\nu ] 
\end{equation}

%
%
%
%
%

\noindent Shifting a function before a convolution is equivalent to shifting after a convolution; therefore, a Doppler shift in $S_\nu$ will cause the quantity $[ S_\nu * \tilde{R}_\nu ]$ to shift by the same amount. The Doppler shifted epoch complex visibility, $B^1_\nu + iC^1_\nu$ relates to an unshifted template complex visibility, $B^0_\nu + iC^0_\nu$, by:

\begin{equation}\label{rv_shift}
B^1_\nu - iC^1_\nu = e^{i2\pi\tau_0\nu} [( B^0_\nu - iC^0_\nu )e^{-i2\pi\tau_0\nu}]_{\nu \rightarrow \nu + \Delta\nu} 
\end{equation}

The change in radial velocity, $\rm \Delta RV$, between the template and epoch can be measured by multiplying the template $B^0_\nu - iC^0_\nu$ by $e^{-i2\pi\tau_0\nu}$, interpolating the product onto a Doppler-shifted $\nu + \Delta\nu$ grid, multiplying that by $e^{i2\pi\tau_0\nu}$, and then comparing the resulting complex visibility to an epoch complex visibility $( B^1_\nu - iC^1_\nu )$.  The visibilities, $V_\nu$, of the template and epoch measurements depend on the quality of interference or contrast in the interferometer in addition to the radial velocity shift, but the phases $\phi_\nu$ only depend on the radial velocity shift.  For this reason, we fit a shifted template to an epoch measurement with $\rm \Delta RV$ as a free parameter, using the error-weighted difference in {\it phases} as the goodness-of-fit statistic.



In configuring TEDI for a particular target, the value of the bulk delay $\tau_0$ is chosen to produce the largest radial velocity signal in $B_\nu$ and $C_\nu$.  This depends on several factors with the rotational broadening of the target's absorption lines playing the largest role.  The rotational broadening relates to the rotation of the star, $\rm V_{rot}$, projected along the angle of the rotation axis with respect to the line of sight, i, such that $\rm \Delta\nu_{rot} = (V_{rot}\sin{i}){\nu \over c}$.  The radial velocity signal is maximized when the periodicity of the interferometer comb, $1/\tau_0$, matches the width of the stellar lines, $\rm \Delta\nu_{rot}$, causing the largest modulation of spectra, $I_{\nu,\tau_0,\Delta\tau}$, with $\Delta\tau$.  Rotational broadening increases linearly with wavenumber, but the period of the comb does not.  This means that each wavelength has a different optimal bulk delay; optimizing the bulk delay across an entire spectrum is typically done by choosing the bulk delay corresponding to the spectral region with the highest line density and highest flux, both of which depend on the spectral type.  For these reasons, the bulk delays available in TEDI were chosen based on performance simulations using rotationally broadened high resolution models of late-type stars provided by Travis Barman using the PHOENIX model \citep[e.g.][]{Fuhrmeister2005}.  We used a bulk delay of 4.6 cm for the results in this paper, corresponding to a projected rotational velocity of 1 to 5 km $\rm s^{-1}$ for a mid-M star.

\subsection{Determining the Interferometer Delay}\label{delay_errors}

In our account of the TEDI data product and $\rm \Delta RV$ measurement, we have relied on accurate knowledge of the bulk delay and phase steps, which together make up the interferometer delay.  However, errors in these quantities will result in errors in the measured change in radial velocity.  An error in the estimation of the bulk delay will appear as a proportional error in $\rm \Delta RV$.  For example, a 1\% error in $\tau_0$ will correspond to a 1\% error in $\rm \Delta RV$.  These errors can be calibrated by observing radial velocity standard stars, and changing the bulk delay to minimize the residuals between the $\rm \Delta RV$ measurements and those expected from the motion of the telescope relative to the Solar System barycenter.

An error in the estimation of the phase steps is much more severe.  If all of the phase steps used to construct the complex visibility are incorrectly estimated by an offset, $\delta (\Delta \tau)$, this will correspond to an offset in the radial velocity change, $\rm \delta (\Delta RV)$, of

\begin{equation}
{\rm \delta (\Delta RV)} = {\delta (\Delta \tau) \over \tau_0} c
\end{equation}

\noindent With a bulk delay of 4.6 cm, a 1 nm offset in the phase steps corresponds to a 7 m $\rm s^{-1}$ error in the measured change in radial velocity.

Ensuring that the interferometer has a specific delay with 1 nm of accuracy for every epoch observation is extremely challenging.  Instead, with TEDI we calibrate the bulk delay and phase steps with emission lines from a ThAr hollow cathode lamp, and use the modulation of the emission lines to correct for delay differences after the data has been collected.

\subsection{Telluric Calibration}

Spectroscopic observations at near-infrared wavelengths must contend with telluric absorption lines introduced by the Earth's atmosphere.  Telluric lines are caused by many species of molecules in the Earth's atmosphere, and the largest contributors at near-infrared wavelengths are $\rm H_2O$, $\rm CO_2$ and $\rm CH_4$.  Telluric lines are numerous across near-infrared wavelengths and will contaminate the stellar complex visibility.  In conventional spectroscopy the standard method for removing telluric lines is to observe a featureless star with a known spectrum, such as a rapidly rotating B star or an A0, Vega-like star, at a similar airmass, and divide that into the target spectrum \citep[e.g.][]{Vacca2003}.  Unfortunately, the complex visibility is not amenable to a simple division, necessitating a more complex and less robust calibration technique.  Simulations of telluric contamination and removal indicate the best method to empirically calibrate telluric lines in TEDI observations is to {\it subtract} the complex visibility of a calibrator from that of the target, normalized to the conventional spectrum $A_\nu$ of the target.  However, incomplete or inaccurate telluric calibration will introduce an error into the measured radial velocity.  Simulations indicate that telluric calibration currently limits TEDI performance, which we address in Section \ref{telluric_simulations}.


\section{TEDI}

A thorough description of the TEDI design and hardware is available elsewhere \citep{Edelstein2010}, so here we briefly summarize only those portions of the instrument's design, components, and history relevant to the current performance.  The TripleSpec spectrograph was commissioned in October of 2007 \citep{Wilson2004, Herter2008}.  TripleSpec is a cross-dispersed long-slit near-infrared echelle spectrograph, dispersing a 1 x 30 arcsecond slit from 1.0 to 2.5 $\rm \mu m$ across 5 orders at resolution 2700 onto 2 quadrants of a Rockwell Scientific\footnote{Now Teledyne Technologies} Hawaii-2 HgCdTe detector.



In December of 2007, we attached an interferometer to TripleSpec to form TEDI.  The original design used a mirror to redirect the telescope beam toward the interferometer, with a pellicle beam splitter used to simultaneously inject ThAr emission light.  The design suffered from non-common path errors between the starlight and ThAr light, described in \citet{Muirhead2010}.  In December of 2009, we removed the interferometer from TripleSpec and upgraded the design to eliminate the non-common path behavior.  The overlapping starlight and ThAr light is now focused onto a fiber before the interferometer, which ensures common path and common delay between the two.  The new design includes two fibers; nodding the target between the fibers nods the target on the TripleSpec detector, allowing for efficient background subtraction.  The upgraded interferometer was attached to TripleSpec in June 2010. 

\subsection{The TEDI Beam Path}\label{tedi_beam_path}

\begin{figure*}[ht]
\begin{center}
\includegraphics[width=5in]{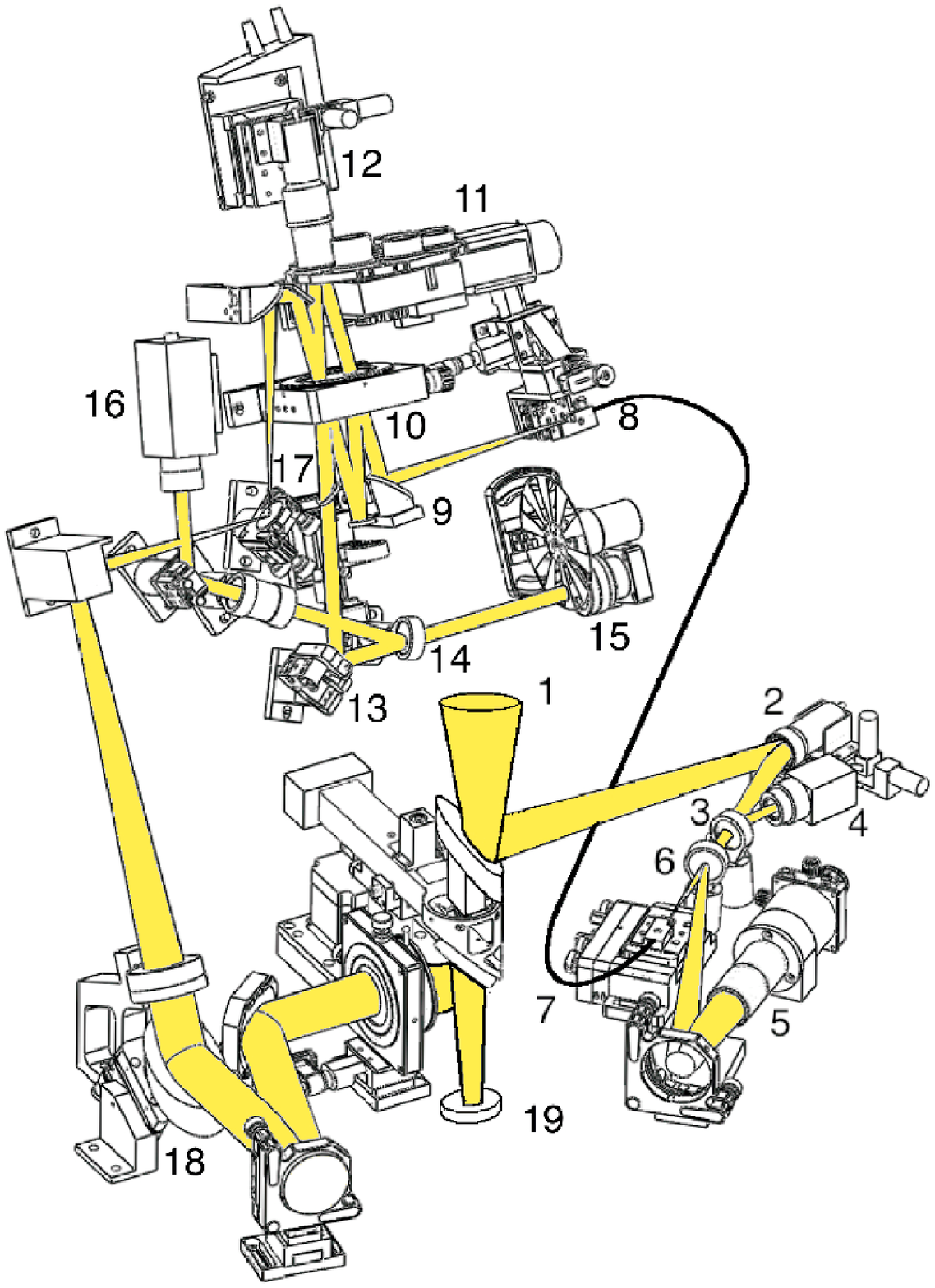}
\end{center}
  \caption[long]{Schematic and beam path of the TEDI Interferometer, described in detail in Section \ref{tedi_beam_path}.  The yellow shaded region corresponds to the beam path.  The pick-off mirror at (1) directs starlight to a guider system which injects the starlight and ThAr calibration light into fibers at (7).  The fibers lead to the interferometer at (8), and eventually to a mirror which redirects the beam down to TripleSpec at (19).  This entire assembly sits above TripleSpec at the Cassegrain focus and fits inside the 39" diameter hole in the 200" primary mirror.}
  \label{tedi_schematic}
  \end{figure*}

Figure \ref{tedi_schematic} depicts the beam path through the interferometer before entering TripleSpec.  First we describe the injection of starlight and ThAr light into the science fibers: (1) A mirror mounted on a removable swing arm deflects the f/16 telescope beam that would otherwise go to TripleSpec. (2) A mirror mounted on a 3-axis, tip/tilt/piston piezo actuator directs the beam towards the two science fibers.  (3) A visible/near-infrared dichroic reflects the visible light to the CCD camera (4) and transmits the near-infrared light to the science fibers.  Once a star has been placed on a science fiber, the CCD camera (4) is used to monitor the location of the star's visible image, and tip/tilt commands are sent to the piezo actuator in (2), ensuring that the star does not drift off of the science fiber.  To introduce the ThAr calibration light a small telescope (5) images a 1 mm fiber core carrying the ThAr emission light to the science fibers at f/16, matched to the telescope beam.  The magnified 4 mm fiber core illuminates both science fibers with ThAr light. (6) A flat window mixes the ThAr light into the science beam by reflection.  (7) The science fibers accept the starlight and ThAr light via glued-on microlenses, which convert the f/16 beam to an f/4 beam. 

Next we describe the interferometer: (8) Starlight and ThAr light exit the fibers via an identical set of microlenses at f/16, with an additional single-mode fiber carrying He-Ne laser light mounted nearby. (9) An off-axis parabola collimates the beams.  (10) The beam splitter directs half of the beam to a fixed mirror, and the other half through one of several etalons (11) to a mirror mounted on a second 3-axis piezo actuator and an adjustable linear stage (12) used in combination with the etalon to introduce the delay.  (13) One interferometer output is directed to another visible/near-infrared dichroic (14), which sends the near-infrared light to a chopper/photodiode system to actively monitor throughput (15).  The second dichroic sends visible light to a second CCD camera (16).  The second CCD camera images the interference pattern produced by the He-Ne laser light.  The pattern is altered by adjusting the tip/tilt/piston piezo actuator in (12).  When the interference pattern contains no fringes, the interferometer is well aligned and produces the highest stellar moir\'e visibilities.  The other interferometer output is directed back toward the TripleSpec window via a fold mirror (17), with an adjustable mirror located at the pupil (18) for positioning the image of the fiber on the TripleSpec slit.  A mirror underneath the removable swing arm (19) redirects the beam down into TripleSpec at f/16.  The entire interferometer assembly sits above TripleSpec at the Cassegrain focus and fits inside the 39" diameter hole in the 200" primary mirror.

\subsection{Throughput}\label{throughput}

TEDI adds many optical surfaces into the beam path prior to entering TripleSpec, 26 surfaces for one interferometer arm and 30 for the other, and uses 50\% of the starlight for monitoring throughput.  This is necessitated by the geometry of adding the interferometer to TripleSpec at its existing Cassegrain mount, but results in significant throughput loss.  With 1 arcsecond seeing, the total throughput of the system--including the atmosphere, telescope, interferometer and TripleSpec--peaks at 1.5\% at 1.68 $\rm \mu m$.  Without the interferometer, the throughput peaks at 30\%.  Focal ratio degradation in the optical fibers also contributes to the lower throughput.  TripleSpec contains a Lyot stop which, when combined with the slit, matches the seeing-limited etendue of the telescope.  Focal ratio degradation in the fibers will effectively increase the etendue of the beam, and introduce irrevocable losses at either the Lyot stop or the slit, depending on the focus.

\subsection{Fiber Scrambling}\label{fiber_scrambling}

The output illumination of a multimode fiber is dependent on various factors including the input illumination geometry. Fiber scrambling methods can be applied to reduce this dependence so that the output fiber illumination is less affected by changes in seeing, telescope pointing or telescope focus. Methods include an optical Òdouble scramblerÓ \citep{Hunter1992,Lovis2006} or a mechanical fiber agitator \citep{Baudrand2001, Ramsey2008}, though the methods are not equivalent and produce different fiber output illumination. TEDI does not use supplemental fiber scrambling methods. 

Partial fiber scrambling is achieved by the microlenses attached to the science fibers, which image the telescope pupil onto the fiber core and act as an image pupil exchanger, similar to a ``double-scrambler'' \citep{Hunter1992}.  However, without identical input illumination, the ThAr and stellar cavity illumination will behave slightly differently.  The ThAr injection is mechanically fixed while the stellar injection will change depending on seeing, guiding and focus fluctuations. This could introduce errors into the calibration of $\tau_0$ and $\Delta\tau$, if the slightly different illumination through the interferometer results in different optical paths.  We measured this effect experimentally by simulating fiber illumination fluctuations and believe it is not a significant source of radial velocity fluctuations.  We discuss the experiment and results in Section \ref{kr_experiment}.


\section{Observational Procedure and Data Analysis}\label{data_analysis}

\begin{figure*}
\begin{center}
\includegraphics[width=6.5in]{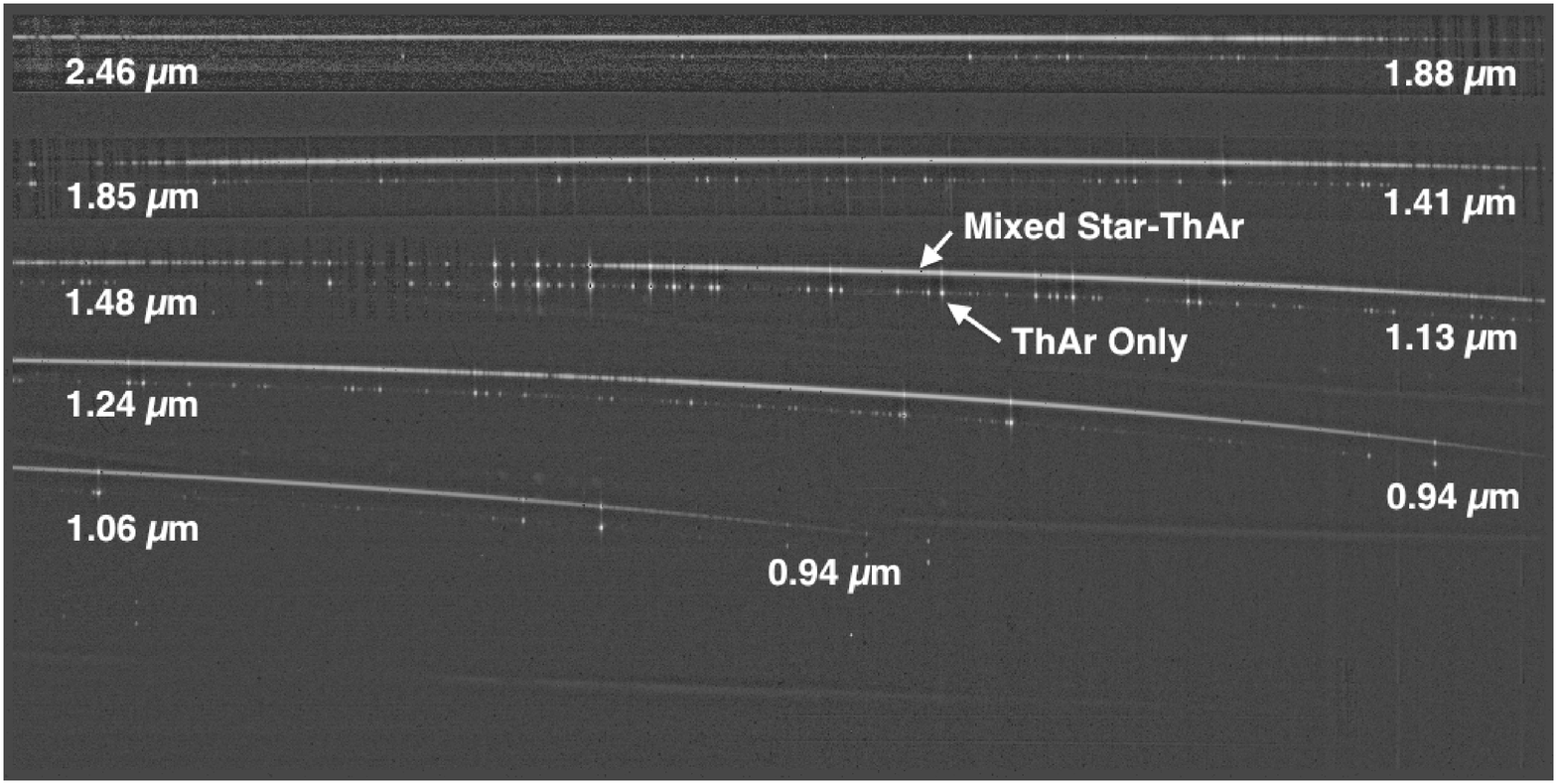}
\end{center}
  \caption{An example TEDI exposure, flat-fielded and dark-subtracted, showing mixed starlight-ThAr lines on the upper fiber, and ThAr lines alone on the lower fiber, annotated to show wavelengths.  Each fiber receives constant ThAr emission light to measure the changes in delay introduced by stepping one of the cavity mirrors with a piezo actuator, with the starlight nodded between the two fibers.  The starlight and ThAr light is cross-dispersed across 5 orders onto 2 quadrants of a Hawaii-2 HgCdTe detector.  Background effects such as thermal emission and OH emission lines from the Earth's atmosphere are removed by subtraction after the spectra have been extracted.}
  \label{figure_exposure}
  \end{figure*}

Figure \ref{figure_exposure} shows a sample TEDI exposure with both mixed star-ThAr light and ThAr light alone.  A single TEDI measurement of the complex visibility in one fiber consists of 20 spectra: 10 of mixed star-ThAr light, and 10 of ThAr alone.  After 10 exposures with the star focused on one fiber, the star is nodded to a second fiber.  Since both fibers receive constant ThAr light, the nodding procedure efficiently provides 10 mixed star-ThAr and 10 ThAr alone spectra on both fibers.  Between each of the 10 exposures we change the delay by $\rm 0.25$ $\rm \mu m$ using a closed-loop Piezo actuator (element \#12 in Figure \ref{tedi_schematic}).  In order to fit the coefficients $A_\nu$, $B_\nu$, and $C_\nu$ in Equation 2, $\Delta\tau$ must Nyquist sample $I_{\nu,\tau_0,\Delta\tau}$ for all $\nu$.   Ten spectra with steps of $\rm 0.25$  $\rm \mu m$ ensures this condition for the full TEDI bandwidth., and we refer to this set of 10 spectra as a {\it phase set}.

Converting 20 exposures, and 40 spectra, into complex visibilities for both fibers involves several steps: spectral extraction, deconvolution of the ThAr lines from the mixed star-ThAr spectra, fitting $A_\nu$, $B_\nu$ and $C_\nu$ to both the ThAr and stellar spectra, accounting for changes in the start delay between template and epoch by referencing $A_\nu$, $B_\nu$ and $C_\nu$ to the same start delay, and subtracting the complex visibility of a telluric calibration star.  With the complex visibilities measured for a template and an epoch, we measure the radial velocity change between the two by fitting a shift as in Equation 9.  We treat each fiber independently, with independent template and epoch measurements.  

\subsection{Spectral Extraction}

Each TripleSpec exposure contains several electronic anomalies which were removed during data processing by modeling the effects and subtracting them out.  An IDL routine for removing the electronic anomalies in TripleSpec exposures is available on the author's website,\footnote{\url{http://astrosun2.astro.cornell.edu/~muirhead/tspec_clean.pro}} and this routine has been used in previous TripleSpec results \citep{Miller2011, Rojas2010, Martinache2009}.  The details of the electronic anomalies are beyond the scope of this paper, but they involve crosstalk of signals between the detector quadrants, and capacitive coupling of signals between the channels within each quadrant.  For a discussion of crosstalk and capacitive coupling in nondestructive arrays, including Hawaii-2 HgCdTe detectors, we refer the reader to Chapter 6.6.4 of \citet{Rieke2002}.


The TripleSpec detector is subject to a variable electronic background, which changes slowly across the detector and slowly with time.  The electronic background is mostly removed by subtracting multiple correlated readouts \citep{Garnett1993}.  However, correlated sampling does not fully remove the electronic background in the detector, which persists and changes with each exposure.  To remove the fluctuating electronic background, we subtract the median of a 20 x 20 pixel box centered around the extraction point.  Each exposure was filtered for cosmic rays and hot pixels using the IDL routine sigma\_filter, available from the IDL Astronomy User's Library \citep{Landsman1993}, which locates pixels significantly brighter than their neighbors.  We flag these pixels as bad, and they are not used in any further analysis.



Individual exposures are flat-fielded and dark subtracted using dome flats and dome darks taken before an observing night.  It is difficult to attain a true estimate of the read noise because TripleSpec does not have the capability to take a ``cold dark,'' wherein a cryogenic mask fully occults the detector.  Instead, we use the dome darks to construct an image of the read noise of the detector, as well as flag any hot or variable pixels.  Dome darks contain significant background emission in K band, and subtracting them from target exposures leaves a negative K band background.  We remove the negative background after the spectra are extracted, however the photon noise from the thermal emission in the dome dark inflates the estimate of the read noise.  

We extract the spectra by (non-optimally) summing pixels across the slit image on the TripleSpec detector.  The TripleSpec slit is tilted on the detector by different amounts at different wavelengths, and summing pixels in this fashion slightly degrades spectral resolution.  The tilt is nowhere more than 10 degrees; as such the resolution loss is less than 3\%.  We chose to sum pixels along the slit to avoid aliasing effects introduced when interpolating the slit onto a rectilinear grid, or interpolating a profile onto the non-rectilinear detector.  We replace bad pixels with the value of a profile fit to the slit image, where the profile is constructed from nearby slit images.  For slit images where more than 40\% of the pixels are flagged, the entire wavelength channel is flagged as bad.  For each extracted wavelength channel, we subtract the median of a 20 x 20 box centered on those pixels to eliminate a fluctuating electronic background in the TripleSpec detector.



\begin{figure}
\begin{center}
\includegraphics[width=6.5in]{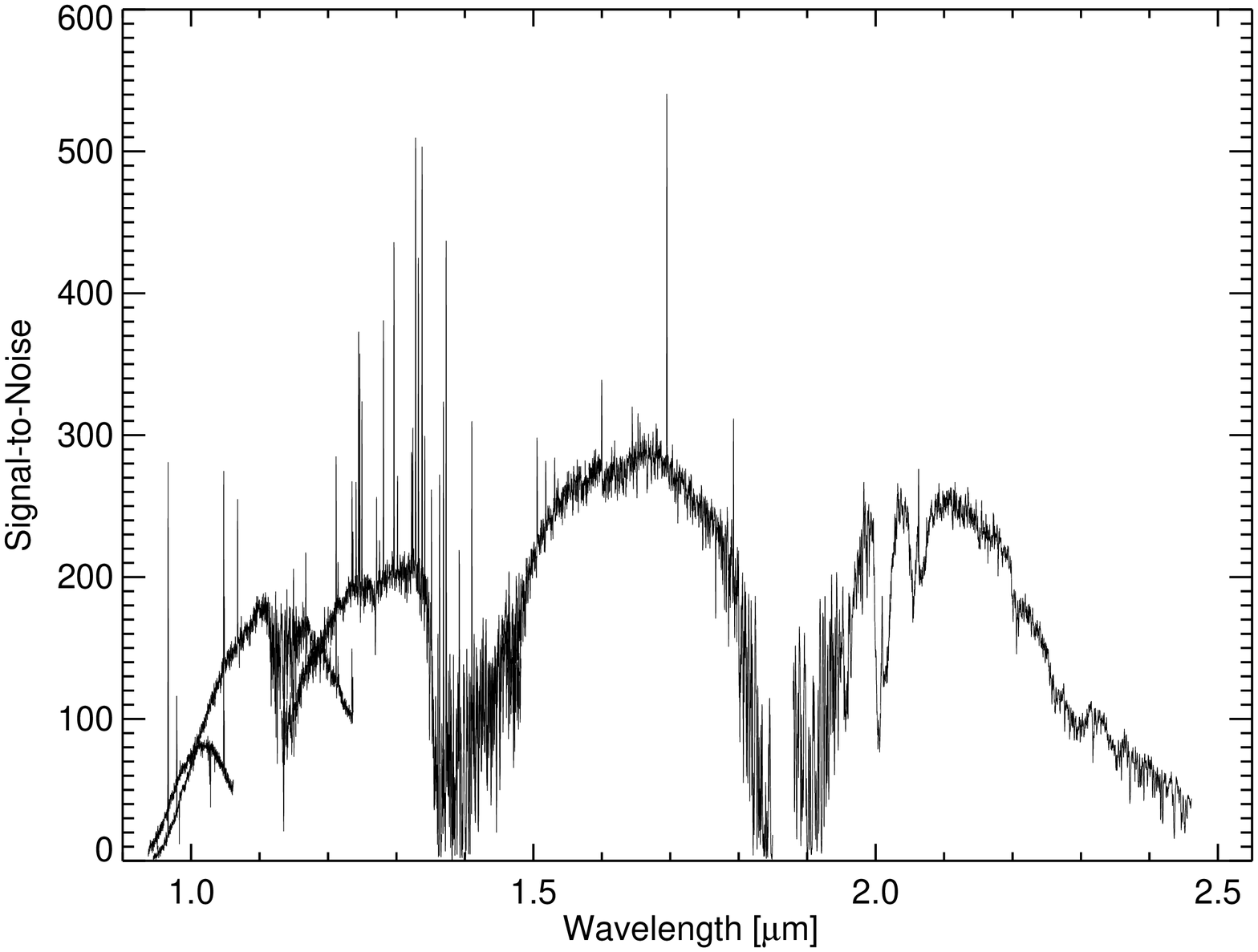}
\end{center}
\caption{A typical TEDI spectrum of mixed stellar (GJ 699) and ThAr light in units of signal-to-noise.  Each of the 5 TripleSpec orders are plotted independently, and they overlap around 1.00 and 1.25 $\rm \mu$m.  Ten spectra, called a ``phase set'', are combined to form the complex visibility (see Figure \ref{moire}).  The extracted spectra are not corrected for instrument efficiency or telluric transmission, since the radial velocity signal is in the moir\'e modulation of the spectra, not the overall shape.  Telluric effects are calibrated out after modeling the moir\'e fringes.  The steep drop in signal-to-noise at 2.2 $\rm \mu$m is due to the combination of higher background and lower transmissivity of the dichroic used for guiding (element 3 in Figure \ref{tedi_schematic}).}
\label{spectra}
\end{figure}

A two-dimensional polynomial wavelength solution for the TripleSpec detector was determined using gas discharge lamps and a slit mask during commissioning of the spectrograph.  TripleSpec has no internal moving parts, which minimizes flexure and changes in the wavelength solution.  However, large changes in the gravity vector of TripleSpec result in illumination shifts of approximately 1 pixel.  After extraction, Gaussian profiles are fit to bright, isolated ThAr lines, and the profile centroids are used to offset to the original wavelength solution to account for flexure.  Figure \ref{spectra} plots an example mixed star-ThAr TEDI spectra, in units of signal-to-noise.

In order to measure radial velocity changes, it is more important that the wavelength solutions of the spectra are the same for every measurement, rather than accurate for every measurement.  This is ensured by cross-correlating the $A_\nu$'s of the ThAr-alone spectra of the epoch measurements onto the $A_\nu$ of the ThAr-alone spectra of the template measurement.  The shift applied to the epoch ThAr spectra is then applied to the corresponding mixed Star-ThAr epoch spectra.  For sub-pixel shifts, the spectra are shifted using spline interpolation.

\subsection{Star/ThAr Separation}

\begin{figure}
\begin{center}
\includegraphics[width=6.5in]{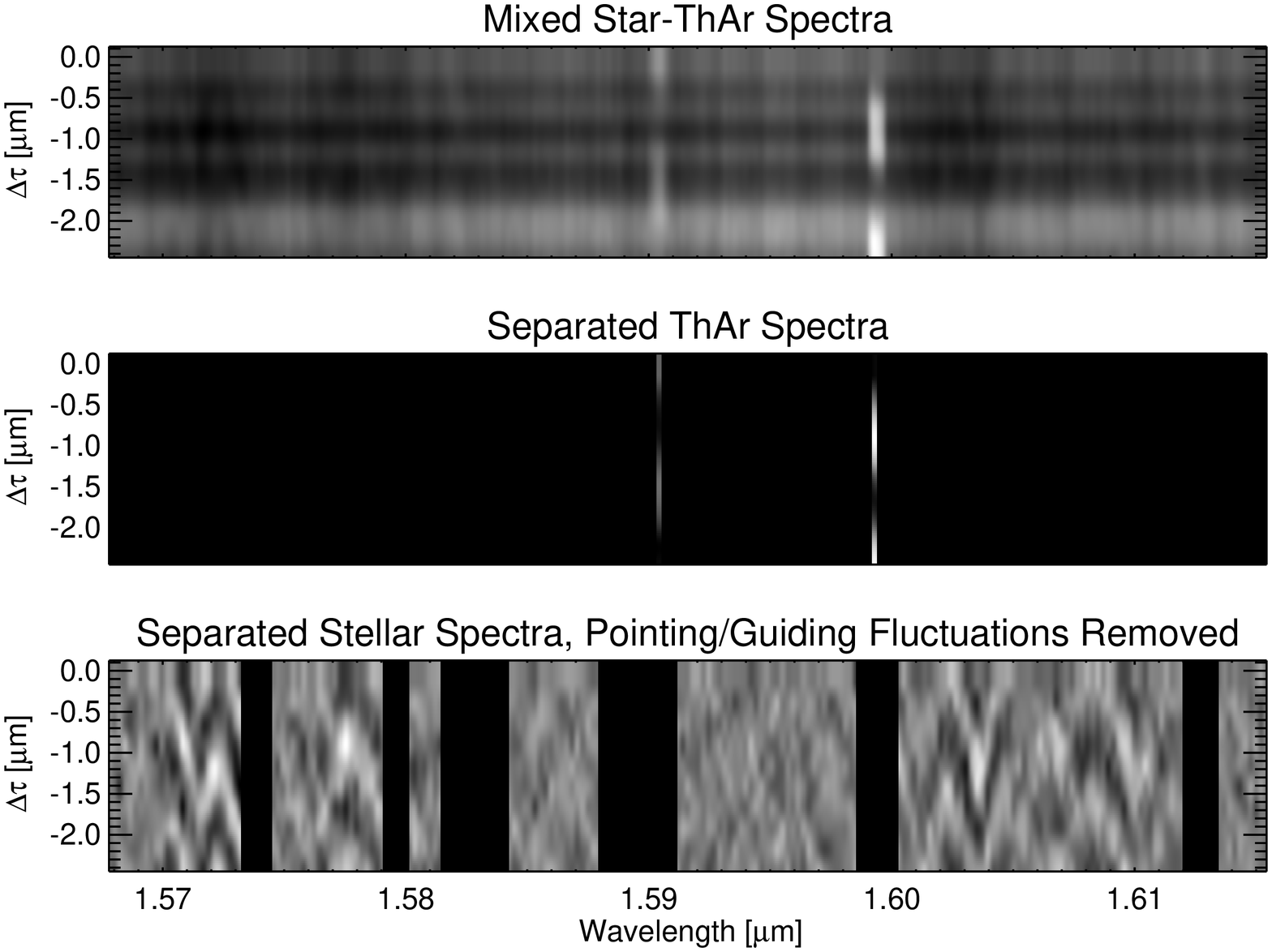}
\end{center}
\caption{TEDI spectra versus wavelength and $\Delta\tau$ for a small region in H-band.  {\it Top:} Mixed stellar (GJ 699) and ThAr light.  The ThAr emission lines dominate the signal, but underneath them lie the stellar spectra.  {\it Middle:} Separated ThAr lines used to accurately and precisely determine the values of $\tau_0$ and $\Delta\tau$.  {\it Bottom:} The stellar spectra separated from the ThAr lines, after being normalized to remove effects from pointing/guiding fluctuations.  The moir\'e fringes are most evident between 1.57 and 1.58 $\rm \mu$m and between 1.60 and 1.61 $\rm \mu$m, and these are due to two $\rm CO_2$ absorption bands in the Earth's atmosphere.  The less distinct fringes between 1.585 and 1.60 are due to the stellar absorption lines, and the phase of the stellar moir\'e fringes is highly sensitive to a Doppler shift.  $A_\nu$, $B_\nu$ and $C_\nu$ are fit to each wavelength channel to model the moir\'e pattern.  Regions with strong ThAr lines or a poor fit to the complex visibility have been blocked out.}
\label{moire}
\end{figure}

To measure the set of phase steps introduced into a mixed star/ThAr phase set, the ThAr lines must be separated from the mixed star/ThAr spectra.  This is done by interpolating the stellar spectrum underneath the ThAr lines and then subtracting that interpolated spectrum.  The locations of the ThAr lines are determined using the ThAr-alone phase set taken immediately before or after the mixed spectra on the same pixels using the nodding scheme.  From here on, we make the distinction between ThAr-alone and ThAr-separated phase sets, the former being the phase set of ThAr alone immediately preceding or following the mixed star-ThAr phase set, and the latter being separated from the mixed star-ThAr spectra.  Figure \ref{moire} plots images of the 10 mixed and separated stellar and ThAr spectra, and shows the resultant moir\'e fringes, to which $A_\nu$, $B_\nu$ and $C_\nu$ are fit.

\subsection{Fitting $A_\nu$, $B_\nu$ and $C_\nu$}
\label{section_fit}




In order to fit the coefficients $A_\nu$, $B_\nu$ and $C_\nu$ to the stellar spectra, we first fit them to the ThAr-alone and ThAr-separated phase sets, which determines the sizes of the phase steps, $\Delta\tau$.  The model for the data is Equation 2, with $A_\nu$, $B_\nu$, $C_\nu$ and 9 of the 10 phase steps, $\Delta\tau$, as the free parameters.  The first phase step is fixed at 0.  The fitting procedure is a joint non-linear Levenberg-Marquardt (LM) fit for the phase steps and a linear fit for $A_\nu$, $B_\nu$ and $C_\nu$ at each wavelength channel.  For each LM iteration of the 9 phase steps, a linear fit of $A_\nu$, $B_\nu$ and $C_\nu$  is performed for each wavelength channel of the phase set.  The best-fit delays, and best fit $A_\nu$, $B_\nu$ and $C_\nu$, are found when the error-weighted $\chi^2$ residuals between the model and data are minimized.  The fitting procedure adjusts the phase steps to ensure that the ThAr lines fluctuate sinusoidally.  This method was chosen over a fully non-linear fit to increase computational efficiency, since there are over 10000 wavelength channels in a given phase set.  A general description of the joint LM-linear fitting method is available in \citet{Wright2009}.


When fitting $A_\nu$, $B_\nu$ and $C_\nu$ to the {\it stellar} spectra, the phase steps are fixed since they are determined by the separated ThAr lines.  Therefore, only the linear portion of the fitting routine is used.  Before fitting the stellar coefficients, the ThAr-alone coefficients are used to create a model of the background at the phase steps of the mixed star-ThAr phase set, and that is subtracted from the mixed star-ThAr spectra.  This removes the thermal background and OH airglow lines present in the spectra, and is equivalent to subtracting a nodded pair of exposures while still accounting for the effect of the interferometer.

The stellar spectra fluctuate due to changes in seeing and telescope guiding errors, which introduce flux variations that are a function of wavelength.  To correct for this, the average shape of each of the 10 spectra is found by convolving each spectrum with a running boxcar mean filter.  The shapes are used to normalize each of the ten spectra and remove these fluctuations.  Various widths for the filter were tested to find that which produced the lowest radial velocity residuals.  If the width of the filter is too large, it will not correct for fast fluctuations versus wavelength, and if it is too narrow it will remove the modulation introduced by the interferometer, which is the signal itself.  We found the width with the lowest residuals to be 11 wavelength channels, or 4 resolution elements.  After the stellar spectra are normalized, $A_\nu$, $B_\nu$ and $C_\nu$ are linearly fit to each wavelength channel.  




\subsection{Referencing Delays}


The LM-linear fitting routine determines the sizes of the phase steps between each spectrum, by ensuring sinusoidal variation of the ThAr lines, but does not determine the difference between the start delay of one set and the start delay of another set.  To reference a complex visibility to a different start delay, we multiply it by an exponential phase change:

\begin{equation}
[B_\nu + i C_\nu] \rightarrow [B_\nu + i C_\nu] e^{2 \pi \delta\tau \nu}
\end{equation}

\noindent where $\delta\tau$ is the change in delay.  To correct for the difference in start delays between two complex visibility measurements, we fit for the $\delta\tau$ which reduces the phase difference between the separated ThAr complex visibilities of a template and epoch measurement.  That $\delta\tau$ is then applied to the epoch stellar complex visibility, such that the template and epoch are referenced to the same start delay.

\subsection{Telluric Calibration}

Telluric calibration is performed by measuring the complex visibility of a rapidly rotating A-type or earlier ``standard'' star near the target, where only the telluric lines will contribute to the complex visibility, and subtracting the telluric complex visibility from that of the target.  The standard complex visibility must be normalized to the target to account for the differences in flux, interferometer visibility and airmass, which will change the visibility as the telluric lines change in depth.  

To normalize the standard star's complex visibility, it is first referenced to the same delay as the target, then divided by $A_\nu$ of the standard and multiplied by $A_\nu$ of the target, which accounts for the change in flux.  Then, a boxcar filter versus wavelength fits a running multiplicative offset to the standard complex visibility to best match the target complex visibility for a given wavelength bin.  A boxcar width of 51 pixels was found to give the lowest radial velocity residuals.  



\subsection{Measuring Radial Velocity Changes}

As mentioned previously, a radial velocity change is measured between a template and an epoch measurement.  The template measurement is identical to an epoch measurement, and is simply chosen as a radial velocity zero-point.  With template and epoch $A_\nu$, $B_\nu$ and $C_\nu$, all referenced to the same delay start point, and the telluric effects at least partially removed, the radial velocity difference $\rm \Delta RV$ can be measured using the procedure outlined in Section \ref{tedi_data_product}.  This requires knowledge of the bulk delay $\tau_0$.  In order to measure the bulk delay, the phase of the complex visibility of the ThAr spectra must be modeled, which requires accurate knowledge of the wavelengths of the ThAr lines.  Using a high-resolution spectrum of a ThAr emission lamp obtained for calibrating the CRIRES spectrograph \citep{Kerber2008}, we directly model the phase of $B_\nu$ and $C_\nu$ using several well-separated ThAr lines in J band, with the bulk delay as the only free parameter.  The delay of the ThAr alone set for the template, which all subsequent sets are referenced to, is chosen for calculating the bulk delay.  As stated in Section \ref{delay_errors}, any residual errors in the bulk delay can be corrected using radial velocity standard stars.

With $\tau_0$ known, the template complex visibility is resampled onto a finer wavelength grid and multiplied by $e^{-i2\pi\tau_0\nu}$, then shifted by $\Delta\nu = {\rm \Delta RV \over c } \nu$ using spline interpolation.  The resulting complex visibility is multiplied by $e^{i2\pi\tau_0\nu}$, returning it to a slowly varying function of wavelength.  This is then resampled onto the original wavelength grid, and the phase $\phi_\nu$, is compared to an epoch measurement.  The $\rm \Delta RV$ which minimizes the $\chi^2$ of the phase difference  between template and epoch, weighted by the formal uncertainty in the phase $\sigma_{\phi_\nu}$, is recorded as the measured change in radial velocity.

\begin{figure}
\begin{center}
\includegraphics[width=6.5in]{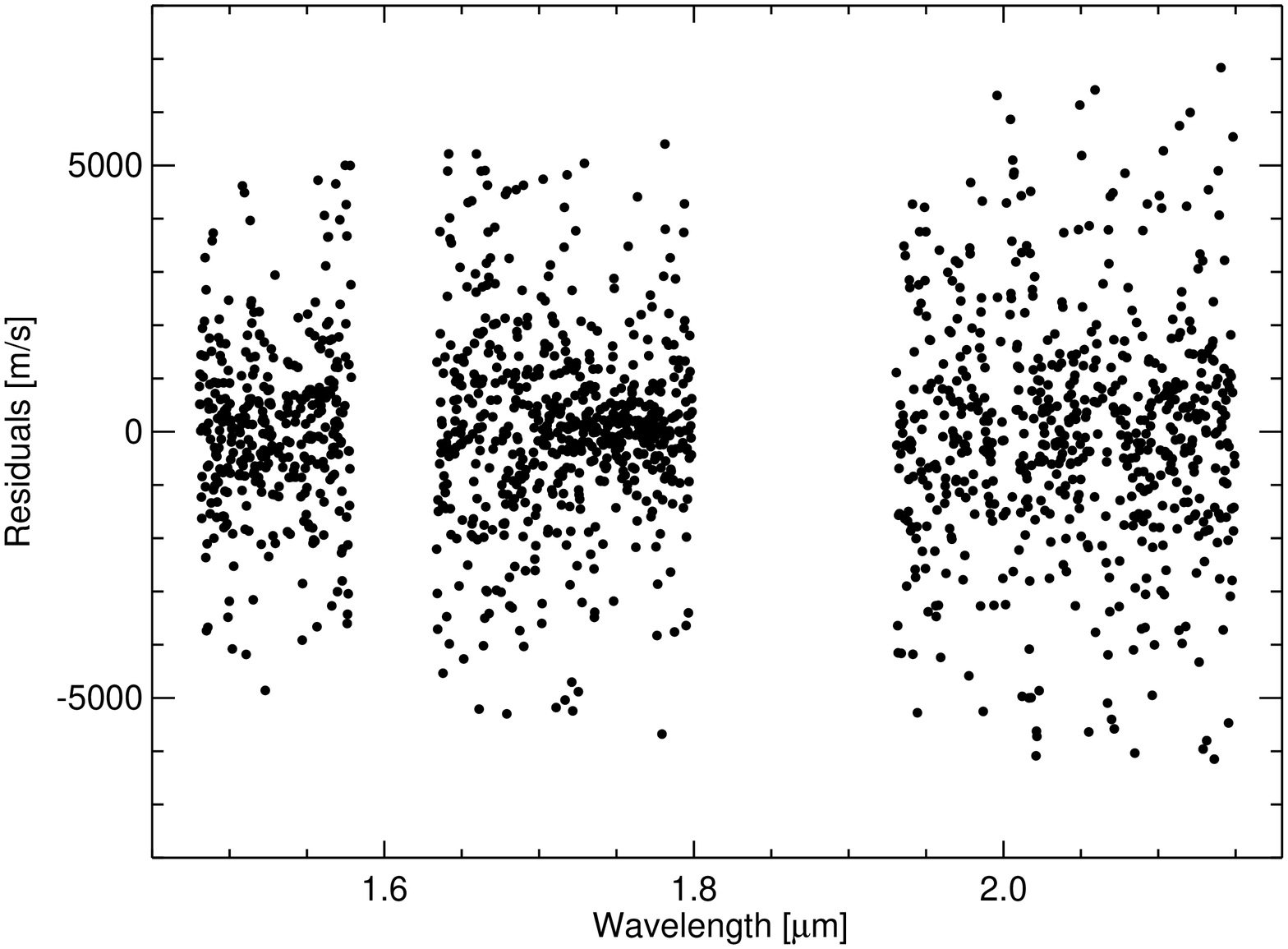}
\end{center}
\caption{Scatter plot of the phase differences for each wavelength channel between the $\rm \Delta RV$-shifted template complex visibility and an epoch complex visibility of GJ 699.  For clarity, the phase differences have been converted to radial velocity differences using Equation \ref{delta_phase}.  The scatter is roughly 1 km $\rm s^{-1}$, but the measured radial velocity depends on the weighted average of {\it all} of the wavelength channels.   The clustering of points indicates spectral regions with narrow stellar features and strong radial velocity signal, with 1.7 to 1.8 $\rm \mu m$ containing the most signal in this bandpass.  We currently limit the bandpass to 1.48 to 2.15 $\rm \mu m$.  The two gaps in coverage are from excessive telluric contamination near 1.62 $\rm \mu m$, which is intentionally ignored, and a gap in TripleSpec's wavelength coverage around 1.95 $\rm \mu m$.  The measured $\rm \Delta RV$ is that which shifts the template to minimize the weighted $\chi^2$ value of these residuals.}
\label{res_wl}
\end{figure}

Instead of using the full TripleSpec bandwidth of 1.00 to 2.46  $\rm \mu m$ for measuring $\rm \Delta RV$, we currently limit our measurements to 1.48 to 2.15 $\rm \mu m$, ignoring regions which have strong telluric absorption.  This bandpass has the highest instrument throughput, strong stellar absorption lines, and was found empirically to deliver lower residuals than when including the full bandwidth.  Figure \ref{res_wl} plots the difference in phase converted to radial velocity between a $\rm \Delta RV$-shifted template complex visibility and a typical epoch complex visibility of GJ 699 for each wavelength channel in this bandwidth.  

\subsection{Formal Uncertainties}

Formal uncertainties for each $\rm \Delta RV$ measurement are found by carrying the per pixel errors in each exposure through the reduction process.  The error in each wavelength channel of each spectrum is calculated by summing in quadrature the read noise and photon noise in each contributing pixel.  Formal uncertainties in $A_\nu$, $B_\nu$ and $C_\nu$ are calculated during the linear portion of the combined linear-LM fitting routine.  The uncertainties in $B_\nu$ and $C_\nu$ are passed to the $\rm \Delta RV$ fitting routine and included to weight the phase difference between the shifted template and epoch.  The formal uncertainty in $\rm \Delta RV$, $\sigma_{\rm \Delta RV}$,  is calculated by converting the formal uncertainty in the phase of the complex visibility to an uncertainty in the radial velocity, and summing the radial velocity uncertainties in each wavelength channel in inverse quadrature:

\begin{equation}
\sigma_{\rm \Delta RV_\nu} = {{ \sigma_{\phi_\nu} c } \over {2 \pi \tau_0 \nu}}
\label{formal_equation1}
\end{equation}

\begin{equation}
\sigma_{\rm \Delta RV} = \sqrt{ 1 \over {\sum{ 1 \over {\sigma_{\rm \Delta RV_\nu}^2}}}}
\label{formal_equation2}
\end{equation}

\begin{figure}
\begin{center}
\includegraphics[width=6.5in]{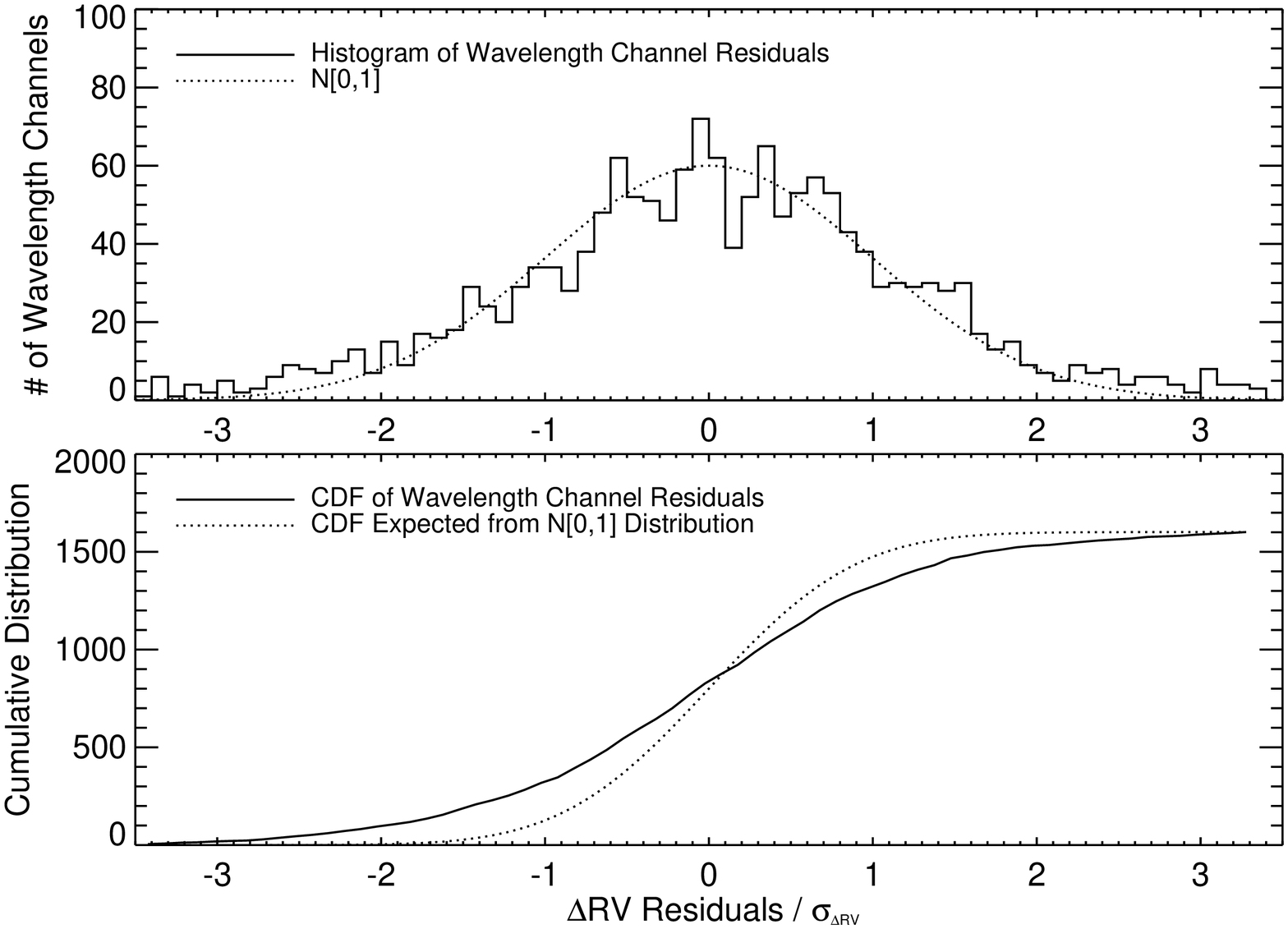}
\end{center}
\caption{{\it Top}: Histogram of the residuals in Figure \ref{res_wl} normalized to their 1$\sigma$ uncertainties, with a normal distribution of mean 0 and standard deviation 1 included (N[0,1]).  {\it Bottom}: Cumulative distribution function (CDF) of the residuals, including that expected from the normal distribution.  The difference indicates significantly more outliers than expected from a Gaussian distribution.  We remove those wavelength channels with the highest 3\% of residuals and repeat the $\rm \Delta RV$ fit to reduce the effect of outliers.}
\label{res_hist}
\end{figure}

Figure \ref{res_hist} plots a histogram of the residuals and the cumulative distribution function of the radial velocities in Figure \ref{res_wl}, normalized to their 1$\sigma$ uncertainties.  The cumulative distribution function indicates significant outliers.  To reduce the effect of wavelength channel outliers, we remove those wavelength channels with the highest 3\% of residuals and repeat the $\rm \Delta RV$ fit.  Currently, errors in the estimates of $\Delta\tau$ and errors in the delay referencing of complex visibilities are not included in the calculation of $\sigma_{\rm \Delta RV}$.  Errors introduced by these effects are presumed to be small, given the high signal-to-noise of the ThAr calibration lines compared to that of starlight.


\section{Observations}

To test the precision of radial velocities measured by TEDI, we conducted a commissioning campaign on GJ 699, also known as Barnard's Star.  GJ 699 is a bright \citep[{\it J}=5.24, {\it H}=4.83, {\it K}=4.52;][]{Skrutskie2006}, slowly rotating  \citep[$\rm V_{rot}\sin{i}<$ 2.5 km $\rm s^{-1}$;][]{Browning2010} M4 dwarf.  Visible-wavelength radial velocity measurements limit potential radial velocity variations to less than 7.2 m $\rm s^{-1}$ \citep{Endl2003}, making it an ideal candidate for testing the radial velocity precision of TEDI.


Each of the 20 exposures which make up two $\rm \Delta RV$ measurements of GJ 699, one for each fiber, was 30 seconds long, achieving a median signal-to-noise per wavelength channel of roughly 250 in H band for each spectrum.  The 20 exposures took approximately 15 minutes of total observing time including exposure time, read time, time used to nod the star between fibers, and time between exposures used to change the delay and communicate between the TEDI control computer and TripleSpec control computer.  We used the highest bulk delay available in TEDI of 4.6 cm, optimized for low projected rotational velocity.  

We took 53 such observations of GJ 699, spread over 11 nights in June, July, August and September of 2010.  That is, 53 $\rm \Delta RV$ measurements with Fiber A, and 53 with Fiber B over 4 observing runs.  For telluric calibration, we observed $\gamma$ Ophiuchi, a V=3.75 A0V star 3.17 degrees away from GJ 699, with a projected rotational velocity of 210 km $\rm s^{-1}$ \citep{Royer2007}.  At such a high projected rotational velocity, any spectral features of the star are significantly broader than the interferometer comb spacing and do not contribute to the measured moir\'e pattern and complex visibility.  We observed $\gamma$ Ophiuchi roughly once for every 4 measurements of GJ 699, or once every hour.

Initially, the $\rm \Delta RV$ residuals correlated proportionally with the expected $\rm \Delta RV$ from the motion of the telescope relative to the barycenter of the Solar System.  This is a consequence of applying an incorrect value for the bulk delay $\tau_0$ in the data reduction procedure, as described in Section \ref{delay_errors}.  To compensate we altered the $\tau_0$ used in the reduction to minimize the $\rm \Delta RV$ residuals, and found a best value of 4.654 cm.  For future measurements, $\rm \Delta RV$ measurements of known radial velocity standard stars will be used to measure the bulk delay for science targets.



\begin{figure*}
\begin{center}
 \includegraphics[width=6.5in]{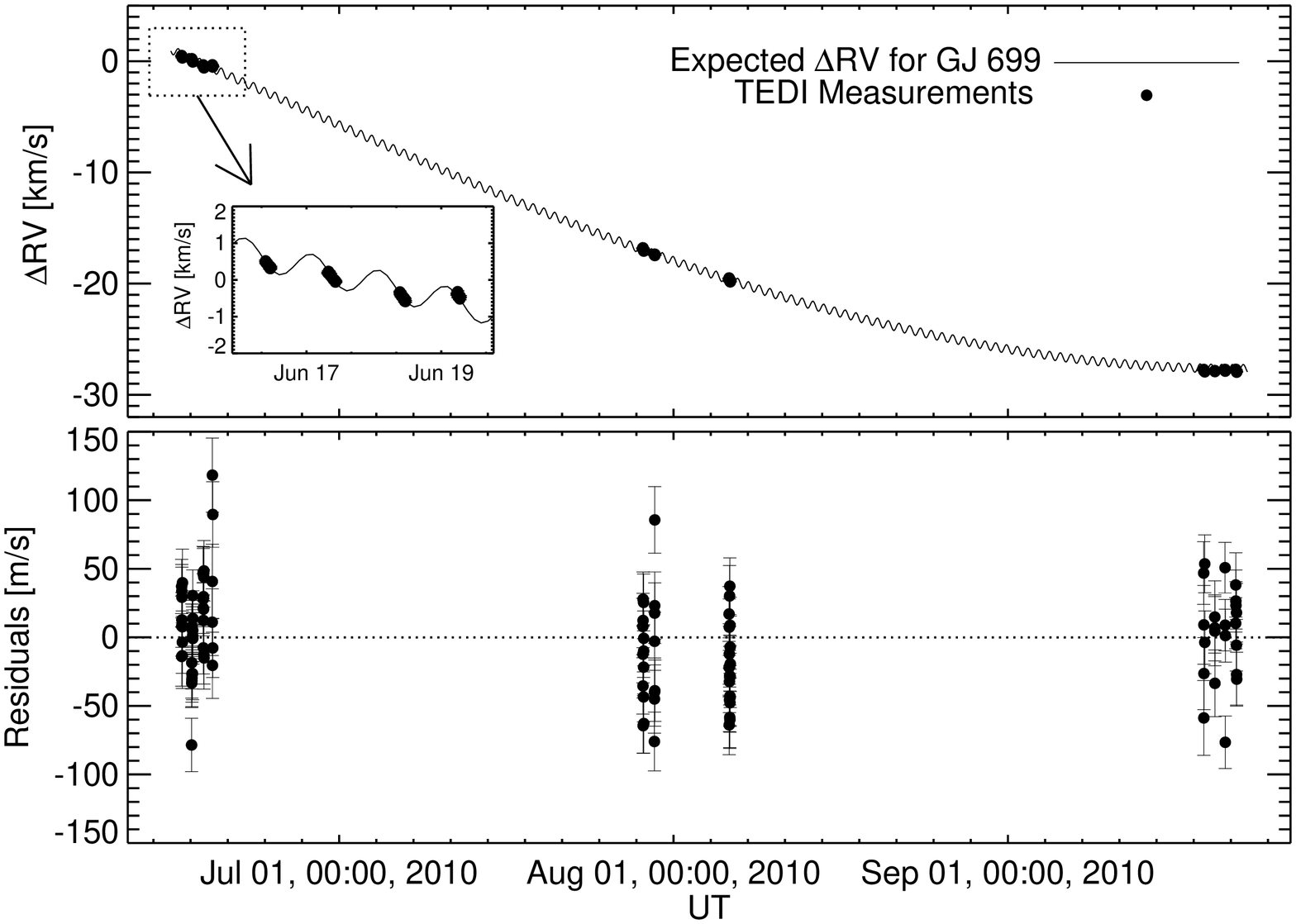}
\end{center}
  \caption{{\it Top}: Expected and measured $\rm \Delta RV$'s of GJ 699 with TEDI.  The expected $\rm \Delta RV$'s are calculated as the motion of the telescope relative to the Solar System barycenter.  This includes a 30 km $\rm s^{-1}$ semi-amplitude component from the Earth's orbit about the Sun, and a 300 m $\rm s^{-1}$ semi-amplitude component from the Earth's rotation.  {\it Top insert}: Detail of 4 nights in June, indicating clear recovery of the Earth's rotation.  {\it Bottom}: Residuals including formal 1$\sigma$ uncertainties.}
\label{rv_resid_ut}
  \end{figure*}
\begin{figure}
\begin{center}
 \includegraphics[width=6.5in]{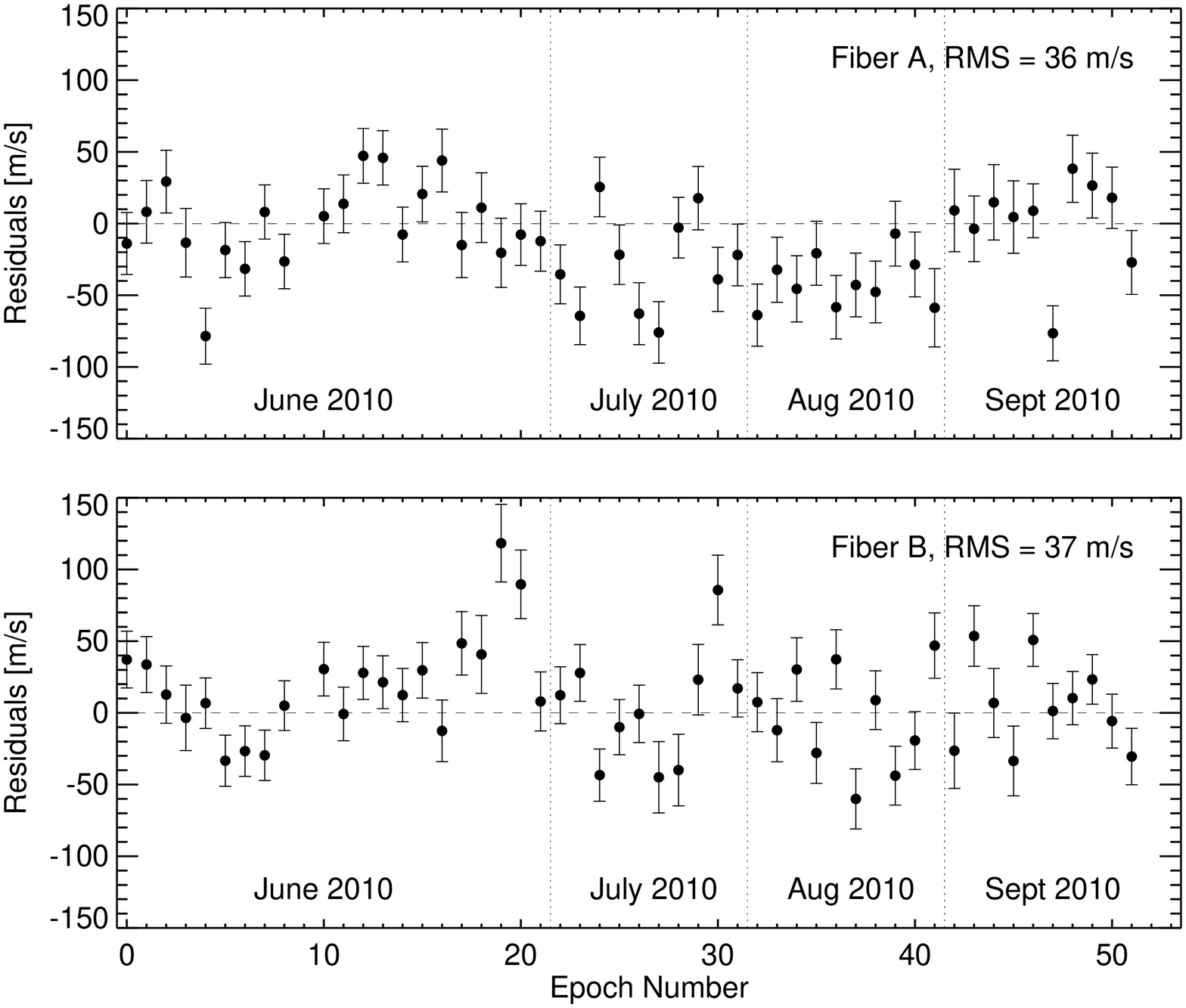}
\end{center}
  \caption{Residuals for each TEDI $\rm \Delta RV$ measurement of GJ 699 with formal 1$\sigma$ uncertainties, equally spaced in order of observation, and grouped by observing run.  The median 1$\sigma$ uncertainty of the measurements is 21 m $\rm s^{-1}$.  {\it Top:} Residuals for Fiber A, with a RMS variation of 36 m $\rm s^{-1}$.  {\it Bot:} Residuals for Fiber B, with a RMS variation of 37 m $\rm s^{-1}$.  The fibers are treated independently, each with its own template measurement taken during the June run.}
\label{rv_resid_epoch}
  \end{figure}
\begin{figure}
\begin{center}
 \includegraphics[width=6.5in]{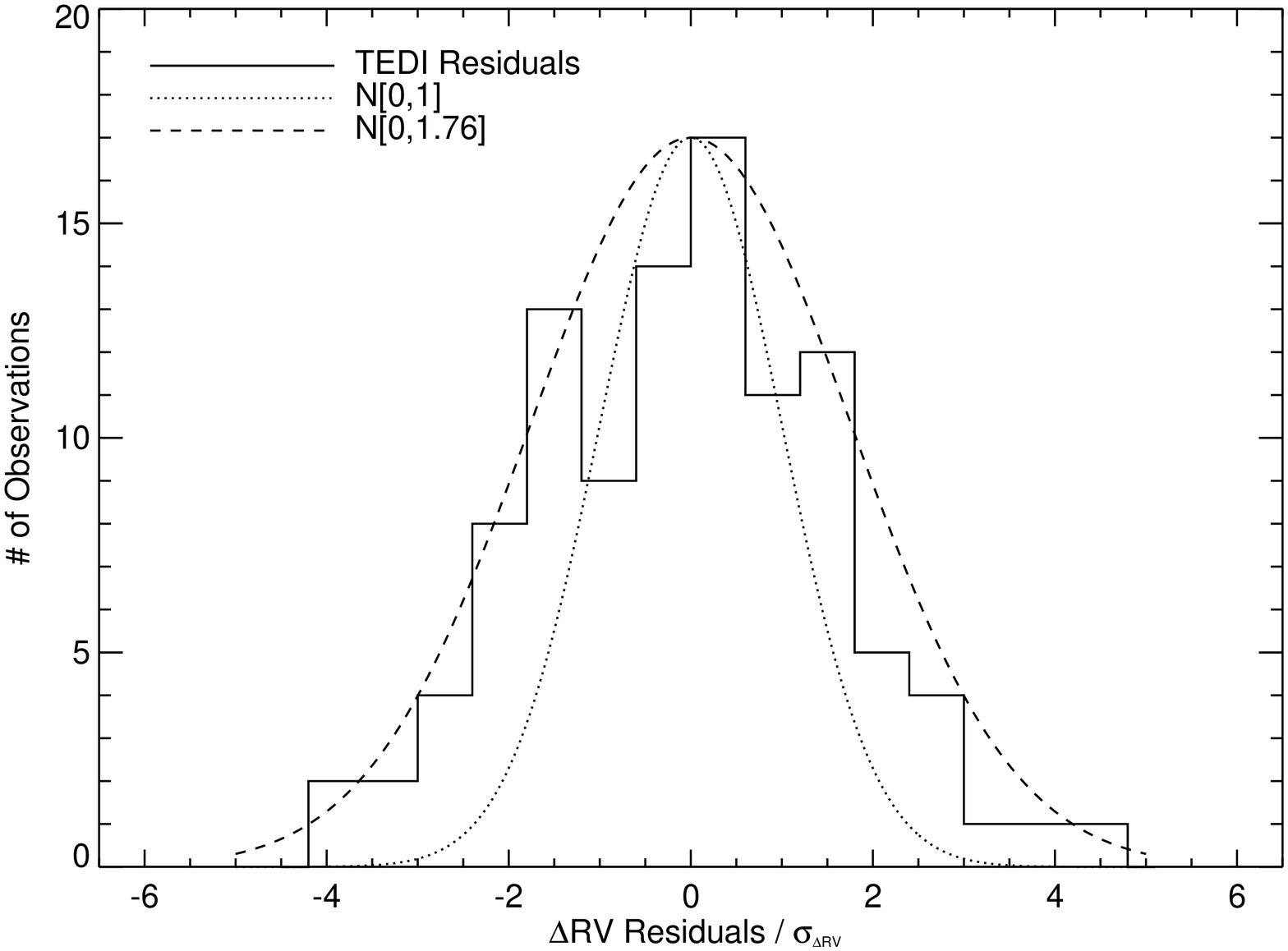}
\end{center}
  \caption{Distribution of the $\rm \Delta RV$ residuals for GJ 699, normalized by their formal 1$\sigma$ uncertainties.  The median formal uncertainty of the measurements is 21 m $\rm s^{-1}$, however the RMS is 37 m $\rm s^{-1}$.  Normal distributions of mean 0 and standard deviations 1 and 1.76 are also shown.  We include a Normal distribution with a standard deviation of 1.76 to account for the difference between the RMS and median formal uncertainty, and this distribution shows significantly better correspondence with the data.}
\label{rv_resid_histogram}
  \end{figure}

Figure \ref{rv_resid_ut} plots expected and measured $\rm \Delta RV$ of GJ 699 over 11 nights in June, July, August and September of 2010, as well as the residuals.  The radial velocities are expected to match the motion of the observatory with respect to the Solar System barycenter.  The fibers are treated independently, with independent template measurements.  The template measurements, used as the zero point for radial velocity changes, were taken on June 17, 2010 and have roughly the same signal-to-noise as the epoch measurements.  Figure \ref{rv_resid_epoch} plots the residuals for each fiber including the 1$\sigma$ formal uncertainties, and Figure \ref{rv_resid_histogram} plots a histogram of residuals, normalized to the formal uncertainties.

\begin{figure}
\begin{center}
 \includegraphics[width=6.5in]{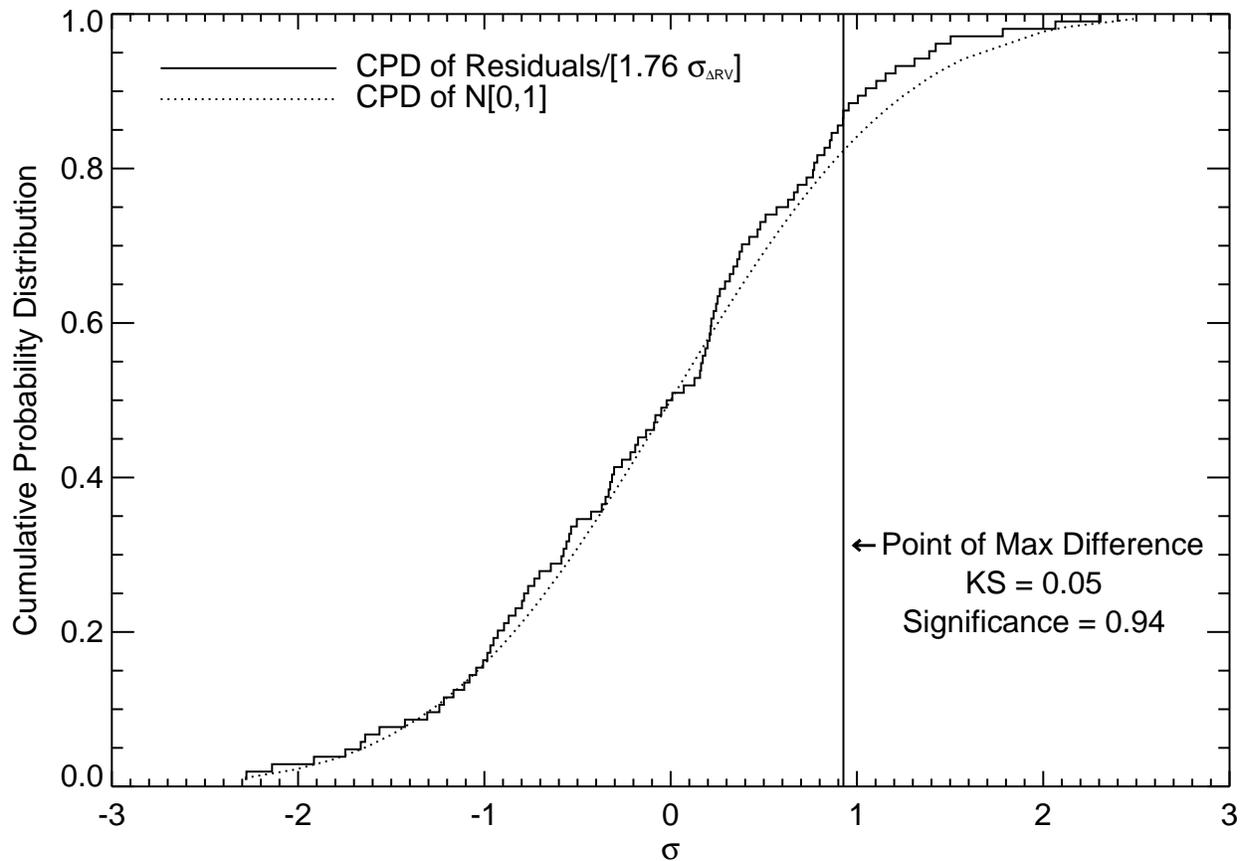}
\end{center}
  \caption{Cumulative probability distribution (CPD) of the $\rm \Delta RV$ residuals for GJ 699, normalized to the formal errors multiplied by 1.76, which is the ratio of the RMS of the residuals to the median formal uncertainty.  This accounts for the difference between the median formal error and the measured RMS of 21 m $\rm s^{-1}$ and 37 m $\rm s^{-1}$, respectively.  The CPD of a Normal distribution with mean 0 and standard deviation of 1 (N[0,1]) is also shown.  The one-sided Kolmogorov-Smirnov statistic is shown, and has a corresponding probability of 0.94, indicating that the distribution of residuals does not differ significantly from a Normal distribution.}
\label{ks}
  \end{figure}

The root-mean-square (RMS) of the residuals is 36 m $\rm s^{-1}$ in Fiber A and 37 m $\rm s^{-1}$ in Fiber B.  This is 1.76 times larger than the median formal uncertainty of 21 m $\rm s^{-1}$.  After inflating the formal uncertainties of the measurements by 1.76, and calculating the one-sided Kolmogorov-Smirnov (KS) statistic between the residuals and a normal distribution of mean 0 and standard deviation 1 (N[0,1]), we found the significance of the KS statistic to be 0.94.  This indicates that the residuals are not significantly different than a normal distribution.  The cumulative probability distributions of the normalized residuals and a N[0,1] distribution are plotted in Figure \ref{ks}.

\begin{figure}
\begin{center}
 \includegraphics[width=6.5in]{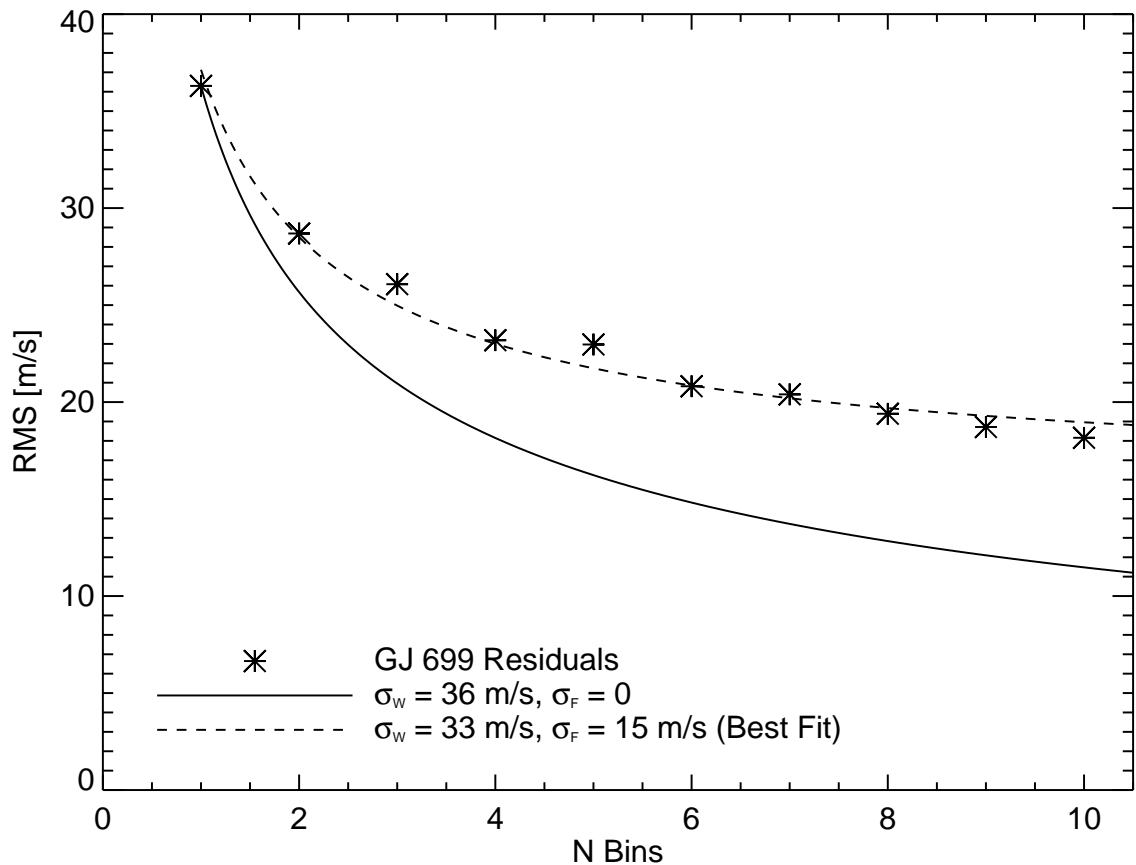}
\end{center}
  \caption{RMS of the $\rm \Delta RV$ residuals for GJ 699 as a function of number of measurements binned together.  Included are plots indicating what is expected from a white noise component $\sigma_W$ and a systematic noise component $\sigma_F$: RMS($N_{\rm Bins}$)=$\sqrt{\sigma_W^2/N_{\rm Bins} + \sigma_F^2}$, where $N_{\rm Bins}$ is the bin size.  The solid line indicates that which would be expected from only white noise matched to the RMS when $N_{\rm Bins}=1$.  The dashed line is fit to the RMS($N_{\rm Bins}$) and indicates a white noise component 33 m $\rm s^{-1}$ which can be reduced by binning, and a systematic noise component of 15 m $\rm s^{-1}$ which cannot.}
  
\label{rms_bin}
  \end{figure}

The individual $\rm \Delta RV$ measurements can be averaged together to reduce the residuals.  Figure \ref{rms_bin} plots the RMS of the residuals versus averaging bin size.  The reduction in RMS is not as fast as would be expected from white noise only.  We model the reduction in RMS as a function of white noise $\sigma_W$ and systematic noise $\sigma_F$, such that $\rm RMS = \sqrt{{\sigma_W}^2/N_{bins} + {\sigma_F}^2}$.  The systematic noise indicates the best precision achievable over these timescales which cannot be reduced by binning measurements.  We calculate a best fit where $\sigma_W$ = 33 m $\rm s^{-1}$, and $\sigma_F$ = 15 m $\rm s^{-1}$.

\begin{figure}
\begin{center}
 \includegraphics[width=6.5in]{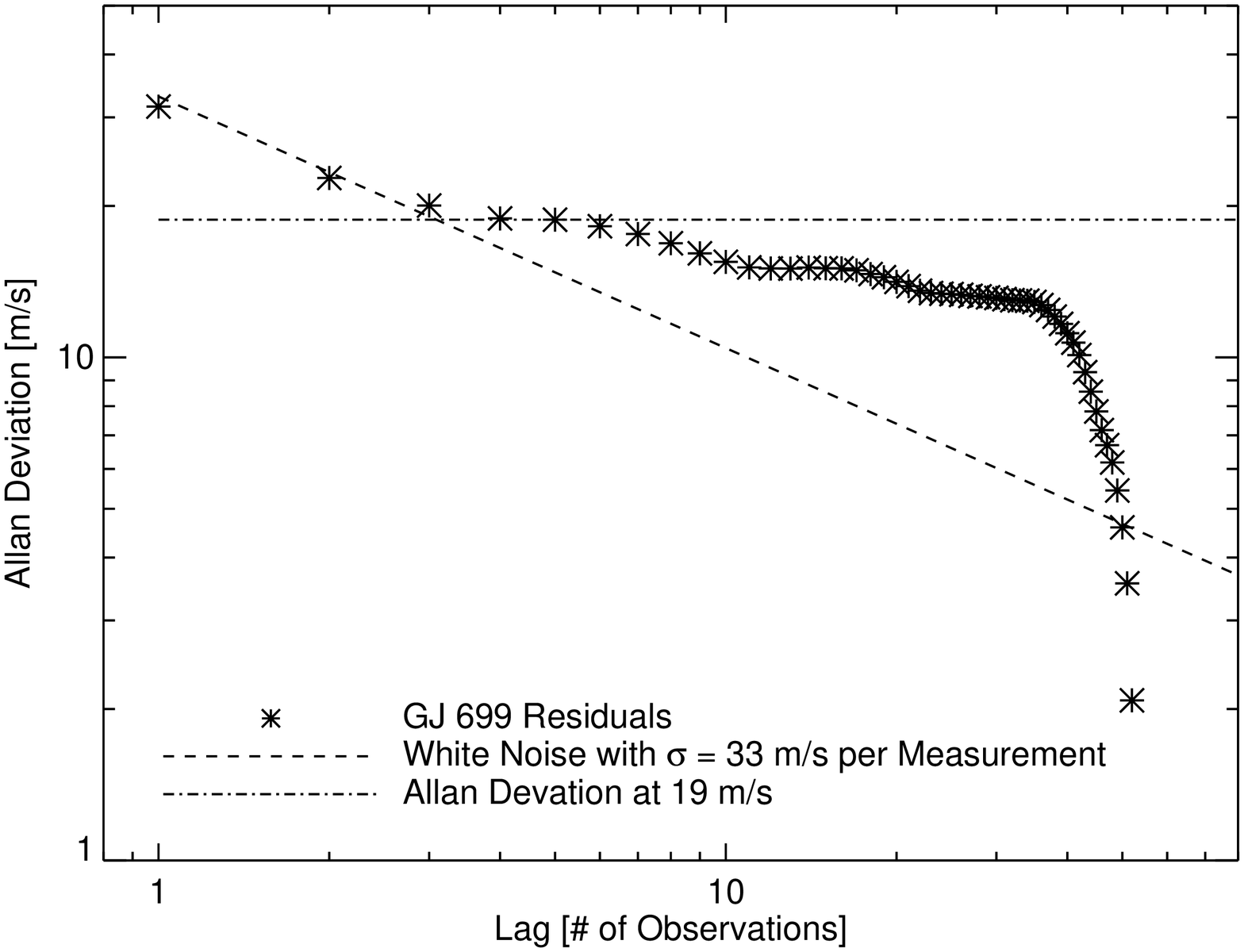}
\end{center}
  \caption{Allan Deviation for TEDI residuals on GJ 699 for each fiber.  TEDI measurements are not equally spaced in time, so the lag used in the calculation is the epoch number, as plotted in Figure \ref{rv_resid_epoch}. Included is the expected contour for purely white noise with a standard deviation of 33 m $\rm s^{-1}$.  The residuals are consistent with white noise until a lag of 4, corresponding to an Allan Deviation of 19 m $\rm s^{-1}$.  This suggests that the non-white contributions to the noise amount to 19 m $\rm s^{-1}$ level, roughly consistent with $\sigma_F$ from Figure \ref{rms_bin}.}
  
  
  
  
  
\label{allan}
  \end{figure}

The residuals in Figure \ref{rv_resid_epoch} appear to have low frequency systematic fluctuations, characteristic of ``red noise'' or 1/f noise.  The epoch measurements are unevenly sampled in time which makes it difficult to make an accurate power spectrum of the noise.  Nevertheless, the noise can be analyzed versus observation number, rather than time.  The Allan deviation, $\sigma_A$, measures the correlation between repeated measurements within a data set \citep[e.g.][]{Thompson2001}, and is given by:

\begin{equation}
\sigma_A^2 \left( k \right) = {\frac{1}{2 \left( N + 1 - 2k \right)}
\sum_{n=0}^{N-2k} \left( \frac{1}{k} \sum_{m=0}^{k-1} y_{m+n} - y_{k+m+n} \right)^2}
\end{equation}

\noindent where $N$ is the total number of measurements and $y_i$ is the residual for measurement $i$.  Figure \ref{allan} plots the Allan deviation vs lag $k$, in units of number of observation number, for the TEDI measurements of GJ 699.  If the residuals were purely white, one would expect the Allan Deviation to decrease with the square root of the lag.  The residuals follow a white noise contour until a lag with an Allan Deviation of 19 m $\rm s^{-1}$.  

The reduction in RMS with bin size and the Allan deviation calculation are both consistent with a noise floor of between 15 and 19 m $\rm s^{-1}$.  It is advantageous to average 3-4 measurements, but beyond that one is rapidly approaching the noise floor, and there are diminishing returns.  In the next section we describe simulations of TEDI data and analysis, and find that incomplete telluric calibration is likely the source of the noise floor.


\section{Telluric Calibration Errors}\label{telluric_simulations}

We expect the largest source of error in the measurements to be due to insufficient removal of telluric contamination in the complex visibilities.  TEDI does not fully resolve the narrowest features of either the stellar or telluric lines.  When unresolved, high resolution features of the stellar and telluric spectrum will mix and alias to lower resolution features which will contaminate the complex visibilities.  The effect of the telluric and stellar mixing on the $\rm \Delta RV$ measurement will depend on stellar type, projected rotational velocity, radial velocity, airmass of the target, and the telluric line strengths.  All of these parameters will change the aliased spectrum and will contaminate the complex visibilities differently.

To investigate this effect, we simulated mixed stellar and telluric spectra and telluric calibration spectra, and carry them them through the analysis procedure.  We used a high-resolution model spectrum of a 3200 K main sequence star provided by Travis Barman, computed from the PHOENIX model atmosphere code \citep[e.g.][]{Fuhrmeister2005}, and a high-resolution model of the telluric transmission calculated for Palomar Observatory by Henry Roe using his custom BFATS code \citep{Roe2002}.  

We convolved the stellar model with a 1 km $\rm s^{-1}$ rotational broadening kernel before shifting it by an input radial velocity, multiplying it by the telluric transmission model, multiplying it by the 10 interferometer transmission combs corresponding to the 10 phase steps in a phase set, and finally convolving it with a Gaussian profile to simulate the TripleSpec line-spread function.  We matched the signal-to-noise of the simulated spectra to that of the GJ 699 data, and added Poisson noise to simulate photon noise.  After this, we fed the spectra into the current data reduction algorithm described in Section \ref{data_analysis}, using the same wavelength regions of 1.48 to 2.15 $\mu$m for measuring the change in radial velocity.  The simulated data does not contain fluctuations due to telescope pointing and guiding, nor ThAr lines, and the delays are assumed to be known exactly.

\begin{figure}
\begin{center}
 \includegraphics[width=6.5in]{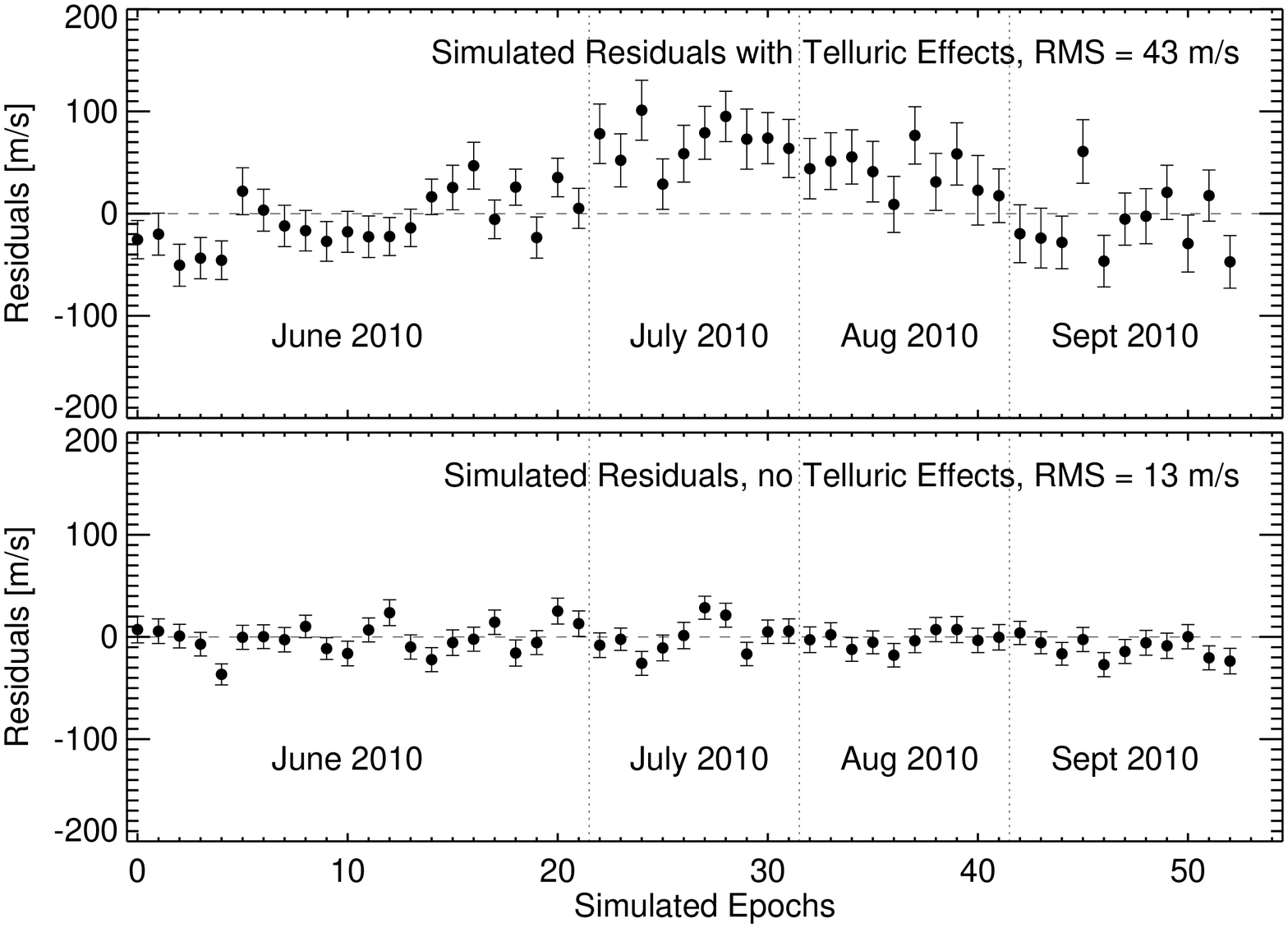}
\end{center}
  \caption{Simulated TEDI measurements of GJ 699 including calculated 1$\sigma$ uncertainties based on photon noise.  {\it Top}: Simulated residuals including telluric contamination and calibration, versus epoch number, showing an RMS of 43 m $\rm s^{-1}$ despite a median 1$\sigma$ uncertainty of 25 m $\rm s^{-1}$.  {\ Bottom:} The same, but without telluric contamination or calibration, and the RMS matches that expected from photon noise of 13 m $\rm s^{-1}$.  The simulations suggest that insufficient calibration of the telluric contamination is currently limiting TEDI performance.  The 1$\sigma$ uncertainties are significantly smaller for the simulations without telluric contamination because the telluric transmission reduces the photon counts at the edges of H and K bands.}

\label{simulated_epochs}
\end{figure}

The expected radial velocities of the GJ 699 measurements were used as inputs, including the bulk radial velocity of the star relative to the Solar System barycenter of -106.8 km $\rm s^{-1}$, previously measured by \citet{Evans1996}.  Figure \ref{simulated_epochs} plots the resulting residuals and 1$\sigma$ uncertainties based on photon noise for simulations with telluric contamination and calibration, and for simulations without any telluric contamination or calibration.  The RMS of the simulated residuals with telluric effects is 43 m $\rm s^{-1}$, roughly a factor of 2 higher than that expected from photon noise, whereas the RMS of the simulated residuals without telluric effects is 13 m $\rm s^{-1}$, matching that expected from photon noise.  The residuals show slow fluctuations, which correlate strongly with the expected radial velocities from the motion of the telescope relative to the Solar System barycenter.  We interpret these errors as due to uncalibrated mixing of stellar and telluric lines. It is not certain whether this is the source of the systematic noise in the TEDI measurements, since the stellar and telluric models are not perfect representations of GJ 699 nor the true telluric transmission, but this analysis suggests it is a plausible explanation.





\subsection{Fiber Illumination Errors}\label{kr_experiment}


TEDI does not currently have a mechanism for supplemental scrambling of the output of the fibers, as discussed in Section \ref{fiber_scrambling}. This could result in slightly different cavity illumination between the ThAr calibration lamp and the starlight, and can potentially introduce errors into the measured $\rm \Delta RV$. The effect was tested by moving an artificial source on the input fiber tip.  To do this, we built a fiber-fed telescope which we mounted on the top of TEDI to imitate a 1 arcsecond ``star'' at nearly the same etendue as the 200" telescope.  We illuminated the fiber with a krypton gas discharge lamp, which has infrared emission lines in H and K bands.  Moving the illumination of the TEDI input fiber by 0.5 arcseconds was found to introduce a 45 m $\rm s^{-1}$ systematic offset in the measured $\rm \Delta RV$.  This could potentially introduce stellar radial velocity offsets if the internal TEDI guider were guiding the starlight to the edge of the fiber core, rather than the center.  However, when the fiber is illuminated 0.5 arcseconds off-axis, the images of the krypton emission lines on the spectrograph detector show a clear double-peaked slit profile, due to an annular shape of the output fiber illumination combined with slit-losses.  The slit profiles of GJ 699 data do not show double-peaked slit profiles.  Since the worst case scenario introduces error on the order of our current RMS performance, 45 m $\rm s^{-1}$ compared to 37 m $\rm s^{-1}$, and we do not see the double-peaked profile, we believe this effect is not currently limiting our performance.



\section{Conclusions and Discussion}

We have demonstrated that the combination of a variable delay interferometer and a resolution 2700 near-infrared spectrograph can achieve better than 37 m $\rm s^{-1}$ of precision with 5 minutes of total integration time per observation on a nearby M dwarf, and that the largest source of error is likely insufficient calibration of telluric lines.  At the current performance level, TEDI is not strongly competitive with current visible-wavelength Doppler surveys of M dwarfs.  Nevertheless, the demonstration of better than 40 m $\rm s^{-1}$ of precision represents an important milestone in the effort to achieve precise radial velocities at infrared wavelengths,  especially considering that the spectrograph involved has a resolution of only 2700.  In order to be competitive with visible-wavelength radial velocity surveys, this technique will need to overcome the effects of insufficient telluric calibration and be implemented with significantly higher throughput.


\subsection{Reducing Telluric Effects}

The effects of insufficient telluric calibration can be addressed by several means.  One solution is to forward model the complex visibilities using high resolution stellar and telluric spectra, either measured by high-resolution spectrographs or modeled by radiative transfer code.  Unfortunately, accurate, high-resolution models of late type stellar spectra are challenging to produce because of the complex molecular opacities which vary significantly with temperature, pressure, and thus photospheric depth.  While telluric transmission models have improved greatly over the last several years, it is unclear whether they are accurate enough for a forward-modeling approach to precise infrared radial velocimetry.
  
Another method of reducing the effect of insufficient telluric calibration is to construct high-resolution spectra by combining many bulk delays from the interferometer.  By combining the complex visibilities from multiple delays, one can reconstruct the initial high resolution spectrum, similar to a Fourier transform spectrograph \citep{Erskine2003b}.  In principal, this method can achieve arbitrary resolution at the expense of survey efficiency, since each delay requires a phase set.  In the case of TEDI, it would require 30 phase sets at different delays to achieve an equivalent resolution of 100000, which is the resolution of CRIRES.  With the stellar and telluric lines fully resolved, or nearly fully resolved, the effects of mixing could be calibrated and reduced.  Although, at this point it isn't clear what resolution is required.


\subsection{Increasing Throughput}

As stated in Section \ref{throughput}, the peak throughput of TEDI is 1.5\% in H band.  Most of the losses occur in the interferometer optical relay system and are due to the sheer number of optical elements, the use of 50\% of the beam for throughput monitoring, and focal-ratio degradation in the fibers.  The product of these losses results in 5\% peak throughput through the interferometer alone.  An optimum implementation of the interferometer could have dramatically higher throughput than TEDI.  A fully fiber-coupled interferometer, placed away from the telescope, would significantly reduce the number of optical elements required and be able to relay {\it both} interferometer outputs to the spectrograph.  With fewer optical elements, and better coupling of starlight into fibers, increasing the throughput of the interferometer up to 25\%, and the total throughput of TEDI to 7.5\%,  is realistic.

\subsection{Technique Development}

Externally dispersed interferometry at near-infrared wavelengths is a propitious technique for measuring precise radial velocities of M dwarfs, but significant challenges remain.  It is important to determine whether the effects of telluric contamination present a fundamental limit to the performance, or whether significant performance gains can be made with better telluric calibration.  If the technique were to efficiently achieve 10 m $\rm s^{-1}$ of RMS precision on an M dwarf, that could dramatically affect the direction of future exoplanet instrumentation.  The scientific returns from precise near-infrared radial velocimetry are rich enough to warrant pursing the last factor of 4 in velocity precision needed to achieve this goal, and attaining this higher precision is promising given the clear avenues remaining for improvement.

\section{Acknowledgements}

We would like to thank the staff at Palomar Observatory for providing significant support toward the implementation and development of TEDI, specifically Steve Kunsman, Kevin Rykoski, Mike Doyle, Bruce Baker, John Henning and Jean Mueller.  We would like to thank the technical staff at the Space Sciences Laboratory who contributed to the TEDI implementation, in particular Mario Marckwordt, Michael Feuerstein, Martin Sirk and Gregory Dalton.  We would like to thank Charles Henderson and Stephen Parshley at Cornell University, who provided technical support and advise in the final few months leading up to the recommissioning of TEDI in 2010.  We would also like to thank Terry Herter at Cornell, the Principal Investigator for TripleSpec, who provided significant support including valuable advice on spectral extraction algorithms, and identification of the TripleSpec detector anomalies.

We would like to thank the members of Conceptual Design Study Team for the Gemini Exoplanet Discovery Instrument (GEDI)--including I. Neil Reid, David Charbonneau, Travis Barman, and Anna Moore--and the conceptual design reviewers--Tom Greene, Tom McMahon, Gordon Walker and Peter Conroy--for their important feedback.  The GEDI Design Study was extremely useful and influential for the TEDI project.  The GEDI Design Study was commissioned as one of the competing PRVS designs and funded by AURA under contract 0084699-GEM00442. The Gemini Observatory is operated by the Association of Universities for Research in Astronomy, Inc., under a cooperative agreement with the NSF on behalf of the Gemini partnership: the National Science Foundation (United States), the Science and Technology Facilities Council (United Kingdom), the National Research Council (Canada), CONICYT (Chile), the Australian Research Council (Australia), Minist\'erio da Ci\^encia e Tecnologia (Brazil), and Ministerio de Ciencia, Tecnolog\'ia e Innovaci\'on Productiva (Argentina).


We would also like to thank the anonymous reviewer of this paper for the thoughtful critique and useful comments.  This work has been supported by the National Science Foundation under Grant No. AST-0505366, AST-096064, AST-0905932, AST-0504874 and by NASA Grant NNX09AB38G.  This work was also supported by NASA Headquarters under the NASA Earth and Space Science Fellowship Program - Grant NNX07AP56H.  The Center for Exoplanets and Habitable Worlds is supported by the Pennsylvania State University, the Eberly College of Science, and the Pennsylvania Space Grant Consortium.  Matthew W. Muterspaugh acknowledges support from the Townes Fellowship Program, Tennessee State University, and the State of Tennessee through its Centers of Excellence program.

\bibliographystyle{/Users/muirhead/Bibtex/apj}
\bibliography{/Users/muirhead/Bibtex/mybib}{}

\end{document}